\documentclass[12pt]{iopart}

\usepackage{bm,enumitem,subcaption}
\usepackage{natbib}
\usepackage{har2nat}   
\setcitestyle{aysep={,}} 
\usepackage{color}
\usepackage{mathrsfs}
\usepackage{graphicx}
\usepackage{epstopdf}
  \expandafter\let\csname equation*\endcsname\relax
  \expandafter\let\csname endequation*\endcsname\relax
\usepackage{amstext}
\usepackage{amsmath} 
\usepackage{mathrsfs}
\usepackage{amssymb}
\usepackage[pagewise, displaymath, mathlines]{lineno}
\usepackage{color}
\usepackage{bm}
\usepackage{booktabs}
\usepackage{xcolor}
\usepackage{float}
\usepackage{breqn}
\usepackage{upgreek}
\usepackage{siunitx}
\usepackage{hyperref}
\hypersetup{colorlinks=true,
	linkcolor=blue,
	citecolor=blue}
\usepackage{booktabs,caption}
\usepackage[flushleft]{threeparttable}
\newcommand{\E}{\mathrm{E}}

\newcommand{\Cov}{\mathrm{Cov}}

\def\spvecA#1;{\if;#1;\else #1\cr \expandafter \spvecA \fi}

\begin{document}

\title[Manuscript for Meas. Sci. and Technol.]{Unscented Kalman filter (UKF) based nonlinear parameter estimation for a turbulent boundary layer: a data assimilation framework}

\author{Zhao Pan$^{1, 2}$, Yang Zhang$^2$, Jonas P. R. Gustavsson$^2$, Jean-Pierre Hickey$^1$, Louis N. Cattafesta III$^{2}$}

\address{$^{1}$University of Waterloo, Department of Mechanical and Mechatronics Engineering, Waterloo, Ontario, N2L 3G1, Canada \\ 
$^{2}$FAMU-FSU College of Engineering, Florida Center for Advanced Aero-Propulsion, Tallahassee, FL, 32310, USA}
\ead{zhao.pan@uwaterloo.ca \&  
lcattafesta@fsu.edu}
\vspace{10pt}
\begin{indented}
\item[] Sept. 2019
\end{indented}

\begin{abstract}
A turbulent boundary layer is a ubiquitous element of fundamental and applied fluid mechanics. Unfortunately, accurate measurements of turbulent boundary layer parameters (e.g., friction velocity $u_\tau$ and wall shear $\tau_w$) are challenging, especially for high speed flows \citep{Smits2010high}. Many direct and/or indirect diagnostic techniques have been developed to measure wall shear stress \citep{vinuesa2017measurement}. However, based on various principles, these techniques generally give different results with {varying} uncertainties. The current study introduces a nonlinear data assimilation framework based on the Unscented Kalman Filter that can fuse information from i)~noisy and discretized measurements from Stereo Particle Image Velocimetry (SPIV), a Preston tube, and a MEMS shear stress sensor, as well as ii)~the uncertainties of the measurements to estimate the parameters of a turbulent boundary layer. A direct numerical simulation of a fully-developed turbulent channel flow is used first to validate the data assimilation algorithm. The algorithm is then applied to experimental boundary layer data at Mach 0.3 obtained in a blowdown wind tunnel facility. Drag coefficients from control volume analysis of the SPIV and wall pressure data and laser interferometer skin friction measurements are used for independent cross-validation. The UKF-based data assimilation algorithm is robust to the uncertain and discretized experimental data and is able to provide accurate estimates of turbulent boundary layer parameters with quantified uncertainty.
\end{abstract}

\vspace{2pc}

\submitto{\MST}

\section{Introduction}
\label{sec:introduction}
\begin{figure}[!htpb]
	\centering
	\includegraphics[width=.7\textwidth]{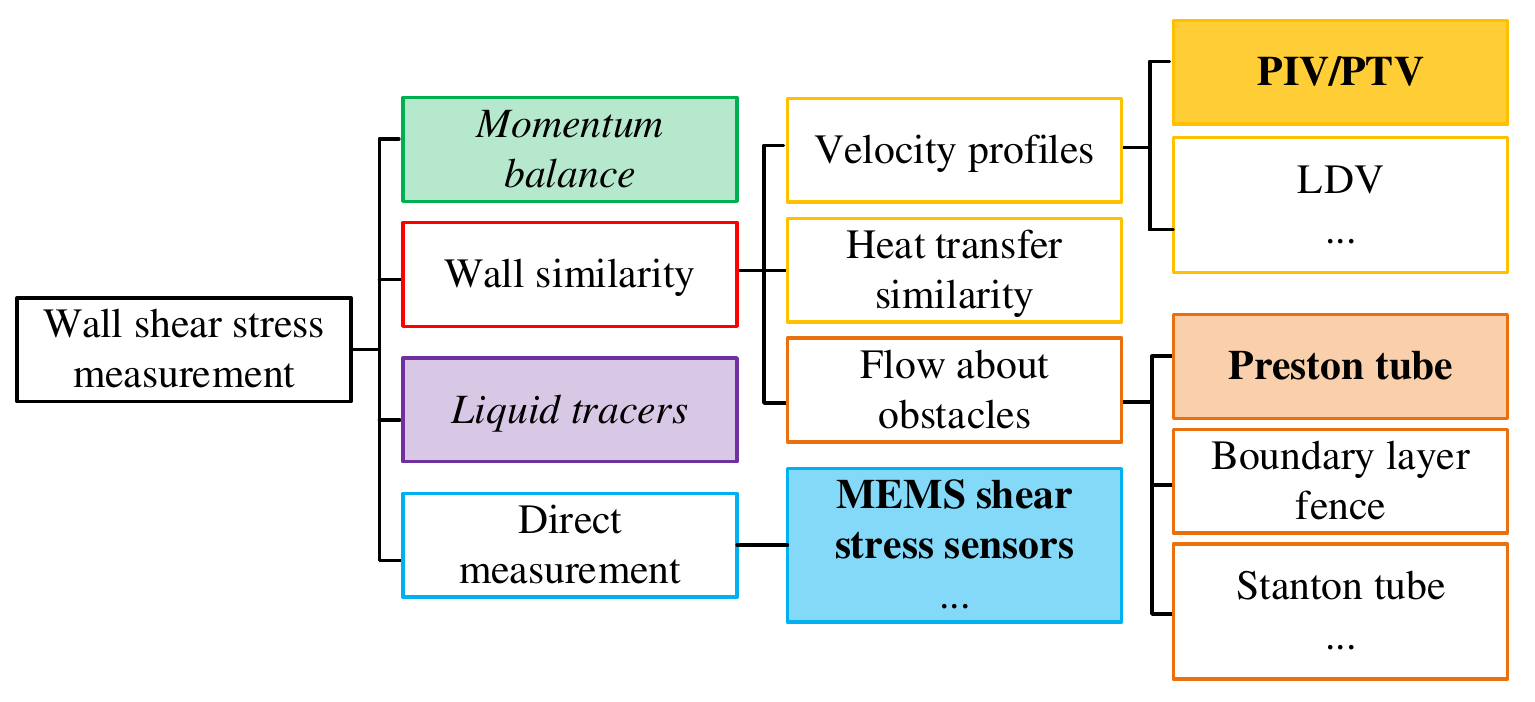}
	\caption{Outline of parameter measurement techniques for turbulent boundary layers (adapted from \cite{winter1979outline}). Techniques employed in the present research are highlighted (bold font for data assimilation and italic font for further validation techniques). }
	\label{fig:tbl_measure_class}
\end{figure}

Despite the advances of computational methods, the experimental determination of the parameters of a turbulent boundary layer (TBL), such as friction velocity ($u_\tau$) and wall shear stress ($\tau_w$), with low uncertainties still remains challenging, especially for high-speed flows \citep{Smits2010high}. Over the years, various diagnostic techniques have been developed and can be classified into four groups as shown in Fig.~\ref{fig:tbl_measure_class} \citep{winter1979outline,haritonidis1989measurement, naughton2002modern,Tinney2018}. These techniques are based on different principles and are characterized by distinct advantages/disadvantages. For example, Particle Image Velocimetry (PIV) is essentially non-intrusive but generally suffers from poor resolution near the wall. The Preston tube measurement is intrusive but relatively inexpensive to carry out.  While modern micromachined noninvasive shear stress sensors can potentially provide direct measurements of $\tau_w$, their rigorous characterization is a topic of current research \citep{mills2017characterization}.

Indeed, high fidelity measurements close to the wall are in general difficult for both intrusive and non-intrusive techniques. Probes (e.g., hot wires and pitot tubes) perturb the flow and are sensitive to their placement \citep{orlu2010near}.  Optical based near-wall diagnostics (e.g., PIV and Particle Tracking Velocimetry (PTV), Laser Doppler Velocimetry (LDV), etc.) are also difficult or even inaccessible due to reflections, sparse particle seeding density, and high flow gradients \citep{kahler2012uncertainty}. To address these problems, one approach is to fit or extrapolate the discretized (and often noisy) data from hot-wire anemometry or PIV to an analytical description of the TBL profile, such as Musker's or Spalding's profile \citep{musker1979explicit,pujals2010forcing, pabon2018characteristics} and the Clauser-Chart \citep{wei2005comment},  or some modified versions of these classic TBL profile descriptions, which are applicable in the viscous sublayer and buffer layer region \citep{kendall2008method,orlu2010near,rodriguez2015robust}.

A nonlinear fitting or regression algorithm commonly looks for a set of parameters in the theoretical expression that minimizes a norm (e.g., $L^1$-~or~$L^2$-norm) of the difference between a fitting function and (noisy) data from experiments. However, it is difficult for such a regression algorithm to directly provide uncertainties of the resulting parameters 
despite the fact that the experimental measurements typically have quantified uncertainties. In addition, a regression method often encounters the following challenges: namely the difficulty in i) overfitting to uncertain experimental data from ‘multi-sensor’ diagnostic setups, where each ‘sensor’ may have different levels and characteristics of uncertainties; ii) assigning appropriate weights or trust to different measurements (e.g., more trust in direct and/or high-fidelity measurements than that from noisy and/or indirect measurements); iii) faithfully taking higher-order information of the measurements into account (e.g., correlated uncertainty of velocity measurements from PIV); and iv) accounting for sensitivity to inaccurate determination of the wall position \citep{orlu2010near}.

To address the above four challenges, we first develop a data assimilation algorithm based on the Unscented Kalman filter (UKF, briefly introduced in \S\ref{sec:ukf}) to estimate the TBL parameters (e.g., $u_\tau$ and $\tau_w$).  The nonlinear UKF is designed to fuse three data sets: i) velocity profiles measured with Stereo-PIV (SPIV), which leads to correlated uncertainties associated with overlap of interrogation windows in the data processing; ii) differential pressure (and its uncertainty) measured by a Preston tube attached to the wall surface; and iii) a micromachined direct wall shear stress sensor and its uncertainty. The algorithm design is described in~\S\ref{sec:data assim}.

Next, we validate the UKF-based data assimilation algorithm in \S\ref{sec:validation} using composite data obtained from a high-fidelity numerical simulation of a channel flow at Mach 0.3, which is contaminated with artificial noise (both correlated and uncorrelated). The UKF-based data assimilation algorithm is then applied to experimental data~(\S\ref{sec:exp}) from a channel flow at Mach $\approx$ 0.3 obtained in a blowdown wind tunnel facility at Florida State University.  Laser interferometer  skin  friction (LISF) and control volume analysis of SPIV and wall pressure data  are  used  for further validation.  Finally, conclusions are offered in~\S\ref{sec:conclusions}.




\section{Unscented Kalman Filter}
\label{sec:ukf}
\begin{figure}[htpb]
	\centering
	\includegraphics[width=.8\textwidth]{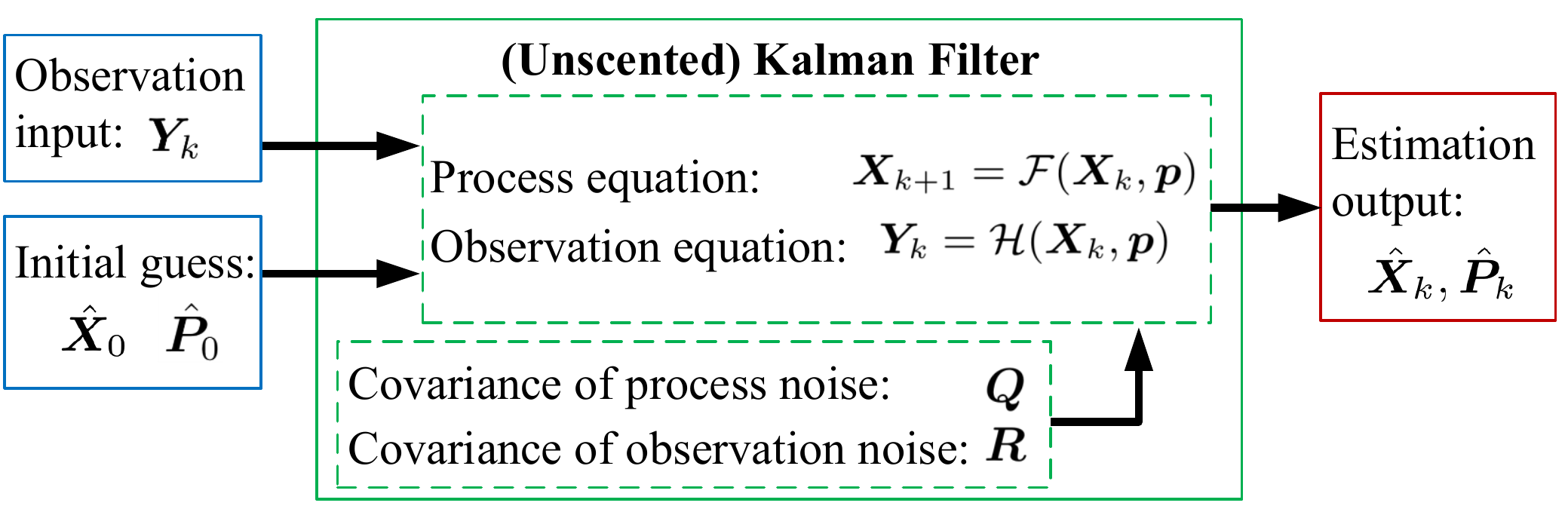}
	\caption{Kalman filter framework for state and/or parameter estimation. A Kalman filter is designed based on i)~the dynamic model of a process ($\mathcal{F}$) and its observable variables ($\mathcal{H}$); and ii)~statistical properties of the process noise and observation noise, which are quantified by covariance matrices $\bm{Q}$ and $\bm{R}$, respectively. Starting with an initial guess ($\hat{\bm{X}}_0$), the filter uses observations ($\bm{Y}_k$) and recursively estimates the unknown state variables or parameters ($\hat{\bm{X}}_k$) and updates the corresponding covariance ($\hat{\bm{P}}_k$), which quantifies the uncertainty of $\hat{\boldsymbol{X}}_k$.}
	\label{fig:ukf}
\end{figure}

The UKF belongs to the family of Kalman filters, which is one of the most common tools for data assimilation, multi-sensor information fusion, and  state/parameter estimation, etc., depending on the context of the field of application. A Kalman filter is commonly designed based on an analytical model of a dynamic process/system $\bm{X}_{k+1}=\mathcal{F}(\bm{X}_{k}, \bm{p})$, where $\bm{X}_k$ is the unknown state vector of the dynamic process at time step~$k$, and the system dynamic model $\mathcal{F}(\cdot)$ is parameterized by $\bm{p}$. The model of the observable variables of the dynamic system is $\bm{Y}_{k}=\mathcal{H}(\bm{X}_{k}, \bm{p})$, where $\bm{Y}_{k}$ is the observation vector containing the variables that can be measured. The mapping between $\bm{X}_k$ and $\bm{Y}_k$ is modeled by $\mathcal{H}(\cdot)$. The inaccuracy of $\mathcal{F}$, which is considered via additive process noise, is modeled by a covariance matrix $\bm{Q}$. Similarly, the observation noise covariance matrix $\bm{R}$  models the uncertainty properties of observations of the system. Initializing the Kalman filter by an initial guess of the unknown state variables ($\hat{\bm{X}}_0$) and given various measurements ($\bm{Y}_k$), the Kalman filter recursively estimates the state variables ($\hat{\bm{X}}_k$) by minimizing the corresponding covariance, $\hat{\bm{P}}_k = \text{cov}(\bm{X}_k - \hat{\bm{X}}_k)$. The diagonal elements of $\hat{\bm{P}}_k$ represent an uncertainty estimate of $\hat{\bm{X}}_k$. The framework of the Kalman filter is illustrated in Fig.~\ref{fig:ukf}, and a detailed derivation and applications of classic Kalman filters are well documented (e.g., \citet{bishop2001introduction}).  The classic Kalman filter is designed for linear and Gaussian processes or systems. For a nonlinear system, the Extended Kalman filter (EKF) is a standard technique that embeds the first-order linearization of the nonlinear system in the Kalman filter framework. Thus, the estimation can be erroneous, especially when the noise is not Gaussian or the process is strongly nonlinear.

The unscented Kalman filter (UKF) is an extension to the Kalman filter, which uses a set of carefully chosen sample points  around the state variables to statistically characterize the dynamics of a nonlinear system. These sample points are called sigma points, which are chosen to have the same mean and covariance as the corresponding state variable  (implementation details can be found in  \citet{wan2000unscented}, \citet{simon2006optimal} and also \ref{appA0}). These sigma points propagate through the nonlinear dynamic system and precisely carry the statistical information of variables (i.e., expected value and covariance). Compared to the EKF, the UKF is superior since it is not restricted to Gaussian statistics and does not linearize the process model. For non-Gaussian statistics, the UKF is accurate to at least the second-order moments (i.e., the expected value and variance) {\cite{wan2000unscented}}. In addition, the UKF {requires significantly less computations} than typical particle methods (e.g., Monte Carlo methods) and it follows the same steps as a Kalman filter for state/parameter estimation (see Fig.~\ref{fig:ukf}). In the present research, we use the UKF to design a state-parameter estimation algorithm to fuse diagnostics from multiple sensors to determine the turbulent boundary layer parameters, such as wall shear stress and friction velocity.

\section{Data assimilation algorithm based on the UKF}
\label{sec:data assim}
A UKF is designed to fuse three different measurements (Stereo-PIV, a Preston tube, and a MEMS floating-element shear stress sensor) to estimate the TBL parameters. 

Velocimetry techniques, such as PIV, are commonly used to measure the mean velocity profile in a TBL. The mean velocity profile in a TBL can be explicitly described by Musker's profile, $u^+ = f(y^+),$ where $u^+ = u/u_\tau$ and $y^+ = y u_\tau/\nu$ ($y$ is the coordinate perpendicular to the wall, $\nu$ is the kinematic viscosity of the fluid, and $u_\tau$ is the friction velocity). In physical units, the velocity profile can be expressed by a modified Musker's profile: 
\begin{equation}
 \tilde{u}(y) = h_1(y, u_\tau, \delta, \Pi, \nu) =
                            u_\tau(u^+_{musker} + u^+_{bump}) \quad  0 \le y \le \delta,
\label{eq:musker}
\end{equation}
where $u^+_{musker}$ is the Musker's TBL profile \cite{musker1979explicit}:
\begin{dmath*}
u^+_{musker} = 5.424 \tan^{-1}\left(\frac{2y^+ - 8.15}{16.7} \right) + \log_{10}\left[\frac{(y^+ + 10.6)^{9.6}}{({y^+}^2 - 8.15y^+ + 86)^2}\right] - 3.52 + 2.44\left[ 6\Pi \left( \frac{y}{\delta} \right)^2 - 4\Pi \left( \frac{y}{\delta} \right)^3 + 
 \left( \frac{y}{\delta} \right)^2\left( 1 - \frac{y}{\delta}\right)
\right], 
\end{dmath*}
and $u^+_{bump}$ is a correction term that offsets the overshot of the original Musker's profile \cite{rodriguez2015robust}:
$$ u^+_{bump} = \frac{\exp \left[-\log^2(y^+/30)\right]}{2.85}.$$ 
Equation \eqref{eq:musker} models the quantitative connection between the measurable portion of the velocity profile, $u(y)$, and variables that cannot be directly measured, such as friction velocity~$u_\tau$ and wake parameter~$\Pi$, and the boundary layer thickness~$\delta$, which can be estimated from measurements. The $(\tilde \cdot)$ above $u$ on the left hand side of \eqref{eq:musker} indicates that $\tilde{u}(y)$ is the known measurement and may be contaminated by measurement error, commonly referred to as `noise': $\tilde u(y) = u(y) + \epsilon_{u_{PIV}},$ where $u(y)$ is the unknown true value of the velocity profile and $\epsilon_{u_{PIV}}$ is the measurement error from PIV. This notation will be adapted hereafter for `noisy' measurements.

Another popular technique to measure skin friction is a Preston tube, which is a pitot probe attached to the wall with a diameter small enough to embed it in the inner part of the boundary layer. The wall shear stress and friction velocity can be derived based on the measured differential pressure between the Preston probe and the local static pressure \citep{head1962preston}: 
\begin{equation}
\log_{10}\left( \frac{\tau_w D^2}{\rho \nu^2}\right) = K_1\log_{10}\left( \frac{\Delta P D^2}{\rho \nu^2}\right) - K_2, 
\label{eq:preston0}
\end{equation}
where $\rho$ is the density of the fluid, $D$ is the outer diameter of the Preston tube, $\Delta P$ is the measured differential pressure, and $K_1 = 0.889$ and $K_2 = 1.400$ are dimensionless constants when $3.7<\log_{10}\left({u_\tau^2 D^2}/{\nu^2}\right)<5.3$ is satisfied \citep{ferriss1965preston}. This nonlinear function \eqref{eq:preston0} maps the relationship between the unknown TBL parameter $\tau_w$ and the experimental observation ($\Delta P$) through wall similarity of flow over an obstacle \citep{winter1979outline}. Rearranging \eqref{eq:preston0} leads to 
\begin{equation}
\Delta \tilde P = h_2(\tau_w, \nu, D, \rho) = \frac{\rho \nu^2}{D^2} 10^{\frac{1}{K_1}\log_{10}\left( \frac{\tau_w D^2}{\rho \nu^2}\right) + \frac{K_2}{K_1}} . 
\label{eq:preston2}
\end{equation}

The wall shear stress can also be directly measured by a miniature floating-element shear stress sensor flush mounted in the wall and can be written as a function of friction velocity: 
\begin{equation}
\tilde \tau_w = h_3(u_\tau, \rho) = \rho u_\tau^2. 
\label{eq:tauw}
\end{equation}

The mean flow profile of a TBL must also satisfy the no-slip boundary condition at the wall ($u|_{y=0} = 0$), which can be expressed as 
\begin{equation}
0  = h_1(y=0).  
\label{eq:u(0)= 0}
\end{equation}
This zero flow velocity at the wall is not necessary to measure in practice, but this extra `data' point at the wall improves the performance of the data assimilation algorithm.

The condition at the edge of the boundary layer, $u(y=\delta) = U_\infty$,  must also be satisfied, where the free stream velocity, $U_\infty$, can be directly determined from velocity profile measurements by PIV. However, $\delta_{0.99}$, corresponding to $u(y=\delta_{0.99}) = 0.99U_\infty$, is often used to approximate the boundary thickness $\delta$. (See \citet{chauhan2009criteria} for more details about the subtle differences between $\delta$ and $\delta_{0.99}$.) Thus, the observation function for the boundary layer thickness is
\begin{equation}
\tilde \delta_{0.99} \approx \delta.
\label{eq:u(delta)= uinf}
\end{equation}

The freestream velocity $U_\infty$ is also measurable using PIV data:
\begin{equation}
\tilde U_{\infty_{PIV}} = U_\infty,
\label{eq:Uinfty}
\end{equation}
where the left hand side ($\tilde U_{\infty_{PIV}}$) is the measurement and the right hand side ($U_{\infty}$) is the unknown true value of the freestream velocity, which is estimated.

Based on \eqref{eq:musker}, \eqref{eq:preston2}~--~\eqref{eq:Uinfty}, we  establish the observation function of the UKF with observation noise $\bm{\epsilon}_o$ as
\begin{equation}
\bm{Y}_k = \mathcal{H}(\bm{X}_k, \bm{p}) + \bm{\epsilon}_o= \begin{pmatrix}
h_1\left(y, u_\tau, \delta, \Pi , \nu \right)\\ 
h_2(u_\tau, \nu, D, \rho)\\ 
h_3(u_\tau, \rho)\\ 
h_1(y=0) \\
\delta\\
U_\infty
\end{pmatrix}_k + 
\begin{pmatrix} 
\epsilon_{u_{PIV}}\\
\epsilon_{PT}\\
\epsilon_{SSS}\\
\epsilon_{0}\\
\epsilon_{\delta_{0.99}}\\
\epsilon_{U_{\infty_{PIV}}}
\end{pmatrix},  
\label{eq:H ukf}
\end{equation}
where 
\begin{equation}
    \bm{Y}_k = \left[\tilde u(y), \Delta \tilde P, \tilde\tau_w, 0, \tilde \delta_{0.99}, \tilde{U}_{\infty_{PIV}} \right]^T,
    \label{eq:Y}
\end{equation}
is the observation vector that is made up by the left hand side of \eqref{eq:musker}, \eqref{eq:preston2}~--~\eqref{eq:Uinfty}.

The setup in \eqref{eq:H ukf} can be interpreted as follows. The observation $\bm{Y}_k$ consists of measurements of unknown variables $\bm{X}_k$ or their functions $\mathcal{H}(\bm{X}_k,\bm{p})$, which are corrupted by unknown error $\bm{\epsilon}_o$ \footnote{ $\bm{\epsilon}_o$ is assumed to be random, as any known bias in the measurements can be directly corrected. Unknown bias can be handled by increasing the measurement uncertainty. A example of this practice can be found in \S~\ref{sec:validation}, where the possible bias of the MEMS shear-stress sensor is accounted greater than its nominal uncertainty.}. Each element in $\bm{\epsilon}_o$ models the estimated measurement noise of the corresponding observation.  For example, $\epsilon_{u_{PIV}}, \epsilon_{PT}, \epsilon_{SSS}$ are the  noise of PIV data, Preston tube measurement, and shear stress sensor measurement, respectively. $\epsilon_0$ represents the small discrepancy between the no slip condition and the modified Masker's profile at the wall (i.e., $u^+(y^+ = 0) = h_1(y=0)/u_\tau = -0.0087 $). $\epsilon_{\delta_{0.99}}$ is the measurement uncertainty of $\delta_{0.99}$, and $\epsilon_{U_{\infty_{PIV}}}$ is the noise of the $U_\infty$ measurement.

The observation noise covariance prescribes the statistics (second moment) of the measurement noise, $\Cov[\bm{\epsilon}_o] = \bm{R}$, and is a \textit{block} diagonal matrix:
\begin{equation}
\bm{R} = \begin{pmatrix} {\bm{\sigma}}_{PIV}^2 & ~  & ~ &~ &~\\
~ & \sigma_{PT}^2 & ~ &~ &~\\
~ & ~ & \sigma_{SSS}^2 &~ &~\\  
~ & ~ & ~ & \sigma_{0}^2 &~\\
~ & ~ & ~ &~ &\sigma_{\delta_{0.99}}^2\\
~ & ~ & ~ &~ & ~&\sigma_{U_{\infty_{PIV}}}^2
\end{pmatrix},
\label{eq:R}
\end{equation}
where ${\bm{\sigma}}^2_{PIV}$ is the covariance matrix of the velocity data measured using PIV, having the same size as the number of PIV measurement locations, ${\sigma}^2_{PT}$ is the variance of pressure measured from Preston tube, and ${\sigma}^2_{SSS}$ is the estimated mean-square error of the shear stress sensor, respectively. PIV processing commonly employs overlapped interrogation windows and leads to correlated uncertainty in the results. Thus, ${\bm{\sigma}}^2_{PIV}$ is typically a banded matrix (e.g., tridiagonal or pentadiagonal, etc., depending on the overlap ratio of the interrogation windows). The diagonal elements (${\sigma}^2_{PIV_{i,i}}$) of ${\bm{\sigma}}^2_{PIV}$ represent uncertainties of PIV measurements at each vector, which can be evaluated by various methods \citep{sciacchitano2015collaborative}, while the off-diagonal elements (${\sigma}_{PIV_{i,i \pm j}}, j = 1, 2, ...$) of ${\bm{\sigma}}_{PIV}$ are nonzero due to the overlapped interrogation windows of PIV measurements. Unfortunately, this higher-order uncertainty information of PIV is rarely evaluated and applied when PIV is used \citep{wieneke2017piv}. Finally, $\sigma^2_{0}$, $\sigma^2_{\delta_{0.99}}$, and $\sigma^2_{U_{\infty_{PIV}}}$ provide uncertainties in the boundary conditions, which are set to be small values, and further explanations are provided in \S\ref{sec:validation}.

The unknown variables in \eqref{eq:H ukf} are contained in $\bm{X}_k$ 
\begin{equation}
\bm{X}_k = [\tau_w, u_\tau, \delta, \Pi, U_\infty]^T, 
    \label{eq:X}
\end{equation}
which is called the state vector in the context of a Kalman filter. The relationship between the variables in $\bm{X}_k$ informs the process model, which is addressed below, and $\bm{p} = [\nu, \rho, D]^T$ are the known parameters of the fluid  and measurement setup.  It is worth noting that some variables appear in both $\bm{X}$ and $\bm{Y}$ but with subtle distinctions. For example, $\tilde \tau_w$ in $\bm{Y}$ is the \textit{known} shear stress measurement from the transducer, which is contaminated by \textit{unknown} error, while $\tau_w$ in $\bm{X}$ is the \textit{unknown} true value of the wall shear stress that is estimated. 
In addition, note that $\delta$ in $\bm{X}$ is the \textit{unknown} thickness of the boundary layer embedded in the modified Musker's profile, which is a mathematical definition of the location where the velocity profile asymptotically reaches~$U_\infty$ \cite{rodriguez2015robust}. Conversely, $\delta_{0.99}$, which is an approximation of $\delta$, is measurable and is employed as an observation in the UKF.

The process model of a Kalman filter describes the relationships between state variables. For example, we have
\begin{equation}
\tau_w = f_1(u_\tau,\rho) = \rho u^2_\tau,
\label{eq:f1_tauw}
\end{equation} 
by definition.
Despite the fact that \eqref{eq:tauw} and \eqref{eq:f1_tauw} take an identical form, note that \eqref{eq:tauw} maps the observation $\tilde{\tau}_w$ between the state variable $u_\tau$, while \eqref{eq:f1_tauw} relates two different state variables.
We also note that the wake parameter $\Pi$ can be calculated as 
\begin{equation}
\Pi \approx f_2(u_\tau,\delta,\nu,U_\infty)=\frac{\kappa}{2}\left[ \frac{U_\infty}{u_\tau} - \frac{1}{\kappa}\log\left( \frac{\delta u_\tau}{\nu} \right) - B \right], 
\label{eq:Pi=}
\end{equation}
where $\kappa=0.41$ is the von Karmnn constant and $B=5.0$ \cite{rona2010generalized}. This equation provides an additional mapping between the unknown TBL parameters in~$\bm{X}_k$.

In the context of state-parameter estimation using a Kalman filter, the unknown parameters can be treated here as pseudo-states of a time-invariant process. The full process model of the UKF is thus formulated as a difference equation of the state vector 
\begin{equation}
\bm{X}_{k+1} = \mathcal{F}(\bm{X}_{k}) + \bm{\epsilon}_p = \begin{pmatrix}
f_1(u_\tau,\rho)\\ 
u_{\tau}\\
\delta\\
f_2(u_\tau,\delta,\nu,U_\infty)\\
U_\infty
\end{pmatrix}_k 
+
\begin{pmatrix}
\epsilon_{\tau_w}\\ 
\epsilon_{u_\tau}\\
\epsilon_\delta\\
\epsilon_\Pi\\
\epsilon_{U_{\infty}}
\end{pmatrix}, 
\label{eq:F ukf}
\end{equation}
where $\bm{\epsilon}_p$ denotes the `process noise' of the system. 
The covariance of process noise of corresponding TBL parameters is set to be a diagonal matrix: $\Cov[\bm{\epsilon}_p] = \bm{Q} = \textbf{diag}\left(\sigma_{\tau_w}^2, \sigma_{u_\tau}^2, \sigma_\delta^2, \sigma_\Pi^2, \sigma_{U_{\infty}}^2 \right).$  Specific values of $\bm{Q}$ are determined by the magnitude and significant digits of $\bm{X}$ and are provided in~\S\ref{sec:validation}.


Equipped with the above setup, the standard routine of the UKF as a state/parameter estimator is implemented (see Fig.~\ref{fig:ukf} and \ref{appA0} for the algorithm). The UKF is designed to provide robust and accurate estimates of the TBL parameters and their corresponding uncertainties based on the measurements from  multiple diagnostic techniques.

\section{Algorithm validation}
\label{sec:validation}
To validate the UKF-based data assimilation algorithm, we constructed synthetic noisy and discrete data sets to mimic the measurements of SPIV, Preston tube, and a wall shear stress sensor based on a high fidelity direct numerical simulation (DNS) of a channel flow at Mach 0.3, which is used as the ground truth. 

A high-fidelity numerical simulation of turbulent channel flow is used to critically assess the data assimilation framework. The scale-resolving simulations are computed using a high-order hybrid LES/DNS code developed by \citet{larsson2013} and \citet{Bermejo2013}. A sixth-order central differencing scheme with high-order artificial dissipation is used to compute the spatial derivatives; the governing equations are advanced in time using a fourth-order, explicit Runge-Kutta scheme. The Reynolds number based on the friction velocity and channel half-height is approximately 750 and the bulk Mach number is 0.3. Doubly periodic boundary conditions are applied to the streamwise and spanwise directions; no-slip is imposed at the wall. The non-dimensional domain size, normalized by the channel half-height, is  $2\pi \times 2 \times \pi$ respectively
in the streamwise ($x$), wall normal ($y$) and spanwise ($z$) direction. The domain is selected to allow an adequate representation of the mean flow and turbulence characteristics \cite{lozano2014}. The grid resolution is $512\times 512\times 340$ in $x$, $y$, and $z$ with a  hyperbolic tangent stretching of the grid in the wall-normal direction; this resolution results in wall unit grid spacing of  $dx^+=9.24$,  $dy^+=$ (0.1 -- 7.5), and  $dz^+ =6.96$. The flow at the wall is resolved to $y^+=0.1$. The dimensional friction velocity of this DNS data is $u_\tau = 4.784$ m/s,  and the half channel height is 2.4~mm. The boundary layer thickness is approximately $\delta_{0.99} = 2.0$~mm.  This new database generated with in-house DNS code is open-access  \href{https://www.frdr-dfdr.ca/repo/handle/doi:10.20383/101.0222}{(doi:~10.20383/101.0222)}  to the public through Federated Research Data Repository (\url{https://www.frdr-dfdr.ca}).

{Table~\ref{tab:2} summarizes the nominal conditions of the DNS and experiments, and more details can be found later in Sec.~\ref{sec:exp}. $Re_\theta$ is the Reynolds number based on momentum thickness, $H$ is the shape factor, $\Pi_c$ is the Cole's wake parameter \cite{coles1956law,jones2001evolution}, $S=U_\infty/u_\tau$ is a skin friction parameter \citep{Perry:2002cg}, $\beta = \frac{\delta^*}{\tau_w}\frac{\mathrm{d}p}{\mathrm{d}x}$ is the Clauser's pressure gradient parameter \cite{francis1954turbulent}, $\mathrm{d}p/\mathrm{d}x$ is the streamwise pressure gradient, and $\delta^*$ is the displacement thickness. 

\begin{table}[h]
		\centering
			\caption{Nominal turbulent boundary layer parameters in the DNS and Experiments.}
	\label{tab:2}
		\begin{threeparttable}
				\begin{tabular}{ccccccc}
			\toprule 
			& $\delta_{0.99}$ [mm] & $Re_\theta$   & $H$ & $\Pi_c$ & $S$ & $\beta$ \\ 
			\midrule
			DNS 		 & 2.07          & 1420    & 1.36     &0.10  &21.9 &0  \\
			Experiments       		 & 3.81        & 2580   & 1.30     &0.43 &23.8 &-0.029 \\
			\bottomrule
				\end{tabular}
		\end{threeparttable} 
\end{table}
}
The mean DNS velocity profile is sampled and cropped to simulate the discrete SPIV measurements that have difficulty resolving the near-wall region. In particular, the DNS data are downsampled to have similar spatial resolution as in the experiments (i.e., 36 vectors in the $55 \le y^+ \le 600$ range which spans over 1.8 mm in physical units). In practice, PIV measurements close to the wall have higher uncertainties primarily due to large velocity gradients. We model these uncertainties by mimicking the uncertainty of the SPIV experiments, specifically by calculating the standard deviation of $N=500$ instantaneous \textit{uncertainty} profiles of SPIV measurements at $i$-th vector ($\sigma^2_{PIV_{i|t}}$~$t= 1,\dots,N$), which is directly exported from a SPIV dataset of a flow at Mach 0.3 (see Fig.~\ref{fig:exp}, and further details are in \S\ref{sec:exp}). The values of the principal diagonal of $\bm{\sigma}_{PIV}^2$ are calculated as 
\begin{equation}
    \sigma^2_{PIV_{i,i}} = \frac{1}{N}\sum^{N}_{t=1}\sigma^2_{PIV_{i|t}}.
    \label{eq:mean TBL error}
\end{equation}

It is worth noting that uncertainty quantification of PIV data, rather than turbulent flow fluctuations, is required to determine the uncertainty of the mean profile. As such, the uncertainty quantification embedded in DaVis 8.4.0 software is used in the present paper \cite{Wieneke:2015dg}.

\begin{figure}[htbp]
	\centering
	\includegraphics[width=\textwidth]{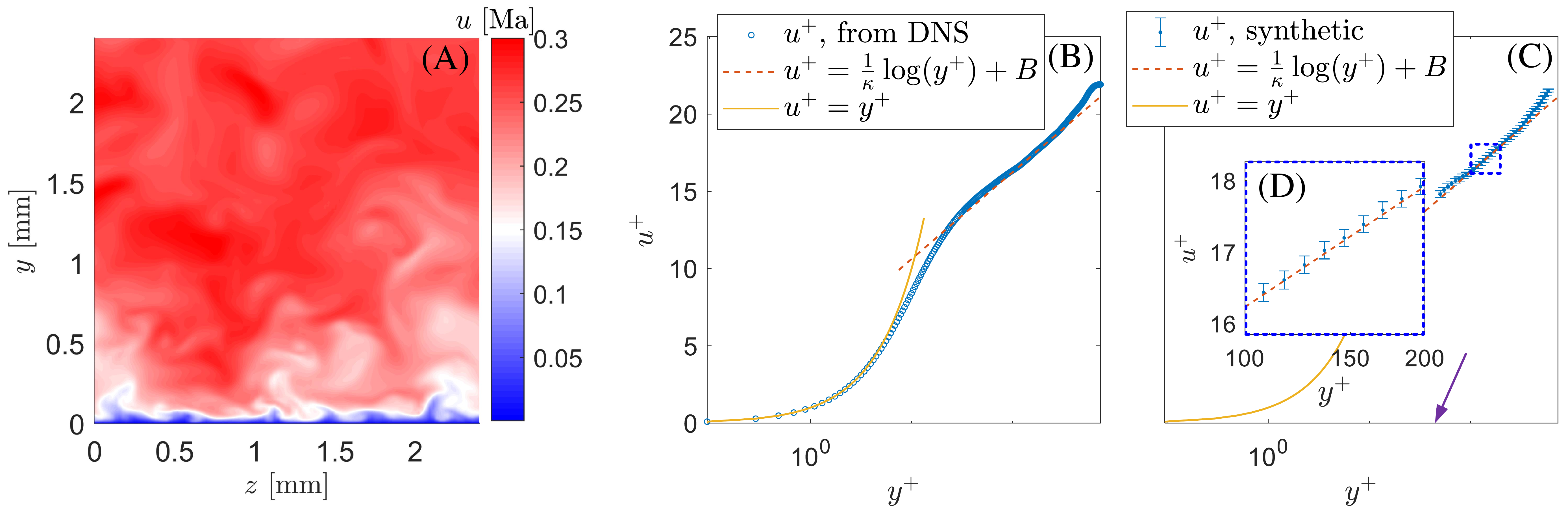}
	\caption{DNS data and DNS-based synthetic data. (A) A slice of the instantaneous velocity field snapshot from the DNS. (B) Mean flow profile (${u}^+$) from the DNS data. The logarithmic law employs constants $\kappa =0.41$ and $B = 5.0$ as reference. (C) Synthetic SPIV data of mean flow profile. A zoomed-in view is shown in (D). Blue dots indicate the noisy and discretized mean flow profile, and the error bars indicate the corresponding uncertainties. The purple arrow indicates the location of the center line of the Preston tube.}
	\label{fig:dns}
\end{figure}

As noted above, the uncertainty of the PIV data is higher near the wall and is also correlated. For example, $\sigma_{PIV}|_{y^+=55} = \sigma_{PIV_{1,1}}= 0.210$ m/s, is larger than that for measurements near the edge of the boundary layer $\sigma_{PIV}|_{y/\delta=0.79} = \sigma_{PIV_{36,36}}= 0.126$ m/s. With $75\%$ overlap of the interrogation windows used in the PIV cross-correlation calculation and a triangular correlation profile (${\sigma_{PIV_{i,i \pm j}}} = (1-j/4){\sigma_{PIV_{i,i}}},~0 \le j \le 4$) \cite{wieneke2017piv,howell2018distribution}, a Cholesky decomposition is used to generate correlated Gaussian noise for the velocity profile ($\epsilon_{u_{PIV_{i}}}$) with the 6 closest vertical neighbors correlated, where each measurement has an assigned variance (i.e., $\epsilon_{u_{PIV_{i}}} \sim \mathcal{N}(0,\bm{\sigma}^2_{PIV})$). Adding this synthetic correlated noise to the mean flow profile from the DNS, we emulate the noisy and discretized synthetic TBL profile measurement from the SPIV data (Fig.~\ref{fig:dns}(C)).

Synthetic pressure measurements of the Preston tube are generated by measuring dynamic pressure of the DNS data at the center line ($y^+ = 46$, indicated by the purple arrow in Fig.~\ref{fig:dns}(C)) of a virtual Preston tube with an outer diameter $D=0.30$~mm. The added noise is Gaussian ($\epsilon_{PT} \sim \mathcal{N}(0,\sigma^2_{PT})$, where $\sigma_{PT} = 0.01\Delta P$ approximates the real Preston probe uncertainty described in \S\ref{sec:exp}). It is worth noting that this synthetic Preston tube estimates $\tau_w$ from \eqref{eq:preston0} that is $\sim7\%$ lower than the value obtained from the DNS. 

Similarly, the shear stress sensor measurement is modeled by adding artificial noise to the exact value of $\tau_w$ calculated from the DNS data. We assumed that the shear stress sensor had high uncertainty due to finite pressure-gradient effects with a non-Gaussian distribution to challenge the data assimilation algorithm: $\epsilon_{SSS} \sim \mathcal{U}(0,\sigma^2_{SSS})$, where $\sigma_{SSS}= {\tau_w}/\sqrt{3},$ meaning that the sensor reading could be anywhere from 0 to 2$\tau_w$, uniformly distributed. $\delta_{0.99}$ is directly extracted from the DNS data by finding the wall distance where $u=0.99U_\infty$, and $U_\infty$ is the velocity at the half-height of the channel.
Gaussian noise are then added to $\delta_{0.99}$, and $U_\infty$ to construct the synthetic measurements: $\epsilon_{\delta_{0.99}} \sim \mathcal{N}(0,\sigma^2_{\delta_{0.99}})$ and $\epsilon_{U_{\infty_{PIV}}} \sim \mathcal{N}(0,\sigma^2_{U_{\infty_{PIV}}}),$ where $\sigma_{\delta_{0.99}} = 0.05\delta$ and $\sigma_{U_{\infty_{PIV}}} \sim 0.02U_\infty$. 
With $\epsilon_{PIV}$, $\epsilon_{PT}$, $\epsilon_{SSS}$, $\epsilon_{\delta_{0.99}}$, and $\epsilon_{U_{\infty_{PIV}}}$ added to the synthetic PIV measurements, Preston tube reading, shear stress sensor measurement and boundary edge measurements from the DNS data, respectively, we have the synthetic noisy experimental data as observation $\bm{Y}$ for input to the UKF. 

In addition, it should be noted that Musker’s profile does not exactly satisfy the no-slip condition (i.e., $u = h_1(y = 0)$ is small but not zero). We therefore set a small value for the `pseudo measurement' uncertainty at the wall: $\sigma_0 = 10^{-3}u_\tau$ ($u_\tau$ is unknown, but an order of magnitude estimate is sufficient here). 
Heuristically, this setup means that the data assimilation algorithm ‘softly’ enforces the no-slip boundary condition.
With the measurement noise covariance ($\bm{\sigma}^2_{PIV}$) and variances (${\sigma}^2_{PT}$ ${\sigma}^2_{SSS}$, $\sigma_{\delta_{0.99}}^2$, and $\sigma_{U_{\infty_{PIV}}}^2$) evaluated, the observation covariance matrix $\bm{R}$ is set up according to \eqref{eq:R}.

Next, we specify the values for $\bm{Q}$ and $\hat{\bm{X}}_0$. In practice, we do not know the exact value of the TBL parameters such as $\tau_w$, $u_\tau$, and $\Pi$ \textit{a priori}. However, anything ranging from order-of-magnitude estimates of the TBL parameters to those obtained from a nonlinear regression (e.g., \citet{orlu2010near}) is adequate to initialize the UKF. For example, $u_\tau \sim \mathcal{O}(10^0)$~m/s for a flow at Mach 0.3, and we retain three significant digits. Thus, $\sigma_{u_\tau} \approx 10^{-2}$~m/s is a reasonable estimate of the process noise for the parameter $u_\tau$. Similarly, we let $\sigma_{\tau_w} \approx 10^{-2}$ and $\sigma_{\Pi} \approx 10^{-2}.$ 
A larger process noise $\sigma_{\delta} \approx 10^{-1}\delta_{0.99}$ is set to account for the difficulty in quantifying the `true' theoretical boundary layer thickness, above which $u(\delta)=U_\infty$ is \textit{exactly} satisfied \cite{chauhan2009criteria}. We choose $\sigma_{U_\infty} \approx 10^{-1}$ for a Mach 0.3 flow, implying that four significant digits are taken for the freestream velocity. Accordingly, let $\bm{Q} = \text{\textbf{diag}}(10^{-4}, 10^{-4}, 4 \times 10^{-8}, 10^{-4}, 10^{-2})$\footnote{Values in $\bm{Q}$ are in the units of $[\text{Pa}^2, \text{m}^2/\text{s}^2, \text{m}^2, -, \text{m}^2/\text{s}^2$].}, and choose an arbitrary initial guess of $\hat{\bm{X}}_0$ such as $\hat{\bm{X}}_0 = (10, 1, 10^{-3}, 10^{-1}, 10^2)$ to run the UKF to give an optimal estimate of the TBL parameters. A large range of initial value, from $0.01\hat{\bm{X}}_0$ to $100\hat{\bm{X}}_0$, have been tested. It has been proven that the algorithm converges to the same estimates, meaning that the algorithm is robust to initial guesses.

\begin{figure}[htbp]
	\centering
	\includegraphics[width=.8\textwidth]{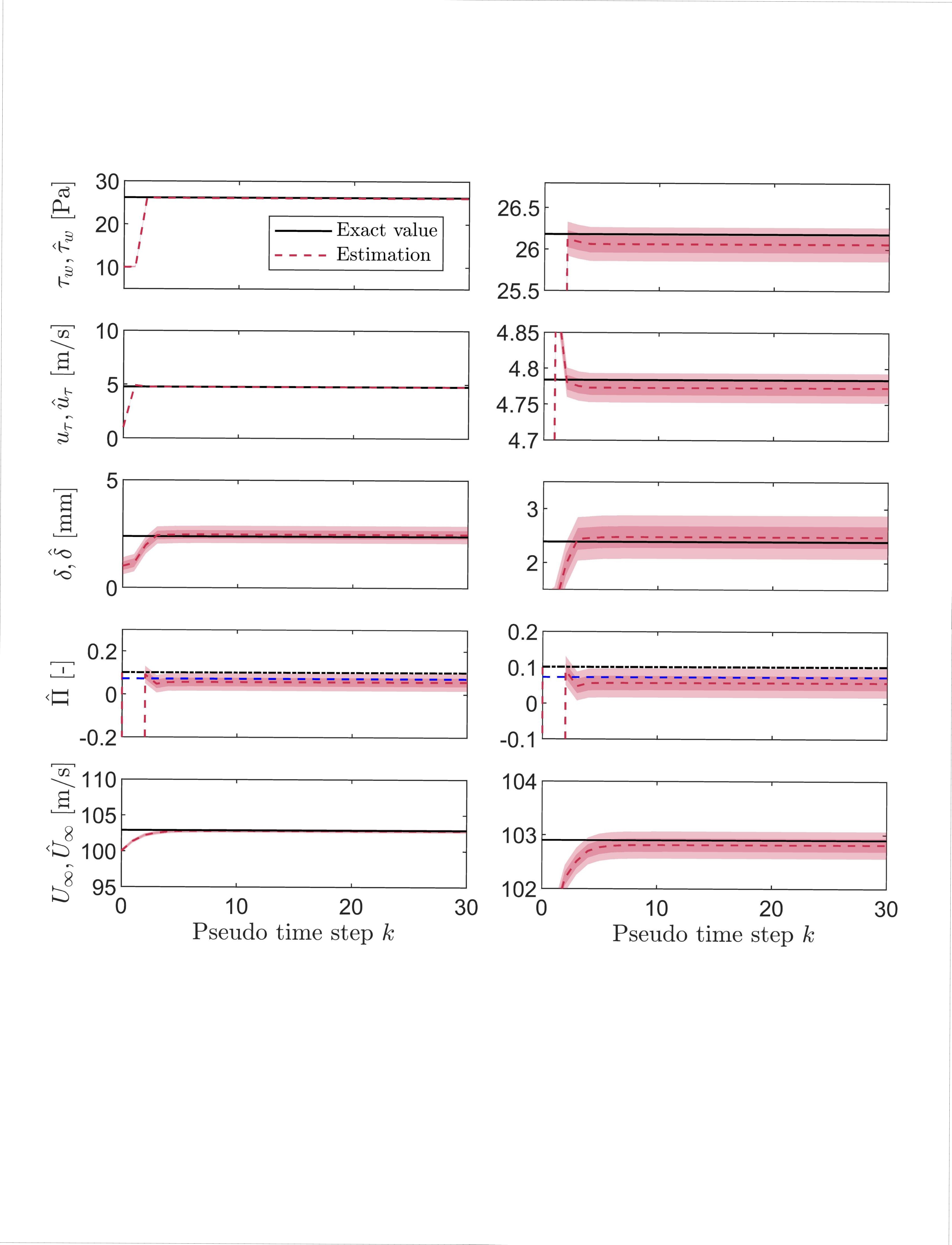}
	\caption{UKF estimation of TBL parameters. The black solid lines indicate the true values from DNS data. The red dashed lines are the estimation output from the UKF. The darker and lighter shades of red indicate the $1\sigma$ and $2\sigma$ uncertainty bands, respectively. The black chain line and the blue dashed line are calculated $\Pi$ using the methods in \citet{jones2001evolution} and \citet{rona2010generalized}, respectively. The corresponding zoomed-in views are shown in the right column.}
	\label{fig:ukfresults}
\end{figure}

A typical data assimilation result is shown in Fig.~\ref{fig:ukfresults}. After several iterations, the UKF converges to optimal estimates of the TBL parameters. The wall shear stress~($\tau_w$) and friction velocity~($u_\tau$) are accurately estimated (red dashed line) with less than $0.5\%$  and $0.1\%$ relative error, respectively. The darker and lighter shades of red patches indicate the $1\sigma$ and $2\sigma$ uncertainty bands of the estimation. The estimated $\hat \Pi$ compares favorably to two different calculations. The black chain line indicates the calculated value of the classic Cole's wake parameter $\Pi_c$ based on the method by \citet{jones2001evolution}. The blue dashed line represents the $\Pi$ calculation based on \eqref{eq:Pi=}, and see \citet{rona2010generalized}.

The UKF-based data assimilation algorithm outputs  $\hat{\bm{X}}_k$ and $\hat{\bm{P}}_k$. $\hat{\bm{X}}_k$ contains the estimated values of the TBL parameters such as $\tau_w$. The diagonal elements of the covariance matrix $\hat{\bm{P}}_k$ are the estimated variances ($\hat{\bm{\sigma}}^2_k$) of the each TBL parameters (e.g., $\hat{\sigma}^2_{\tau_w}$ for $\tau_w$). This $\hat{\sigma}_{\tau_w}$ can be considered as an uncertainty quantification of the estimated TBL parameters. We expect that the estimates from UKF (e.g., $\hat{\tau}_w \pm 2\hat{\sigma}_{\tau_w}$) statistically encompass the true values of the TBL parameters (e.g., $\tau_w$) with 95\% confidence. 

To test the robustness of the UKF-based data assimilation algorithm, we introduce independent random noise to the aforementioned simulated PIV, Preston tube, and shear stress sensor measurements, and run the data assimilation algorithm. This process is then repeated 5,000 times, which is essentially a Monte Carlo (MC) analysis. As the simulated measurements are contaminated by different random noise, each run of the data assimilation may output different results. We shall keep in mind that the UKF treats all TBL parameter as random variables, and each single run of data assimilation returns an expected value ($\hat{\bm{X}}$) and its uncertainty ($\hat{\bm{\sigma}}$) already. The discrepancy between a true value ($\bm{X}$) and estimate ($\hat{\bm{X}}$) is error, and the distribution or variance of this error is interpreted as the uncertainty of the estimate. The averaged error ($\E[\hat{\bm{X}} - \bm{X}]$) from many simulations can thus be interpreted as an expected error from estimation. Similarly the averaged estimated variance ($\E[\hat{\bm{\sigma}}]$) can be interpreted as a measure of the expected uncertainty of many tests. We use this averaged error and uncertainty of the UKF outputs to assess the performance of the data assimilation algorithm when the data are perturbed.

\begin{figure}[h!]
	\centering
	\includegraphics[width=.5\textwidth]{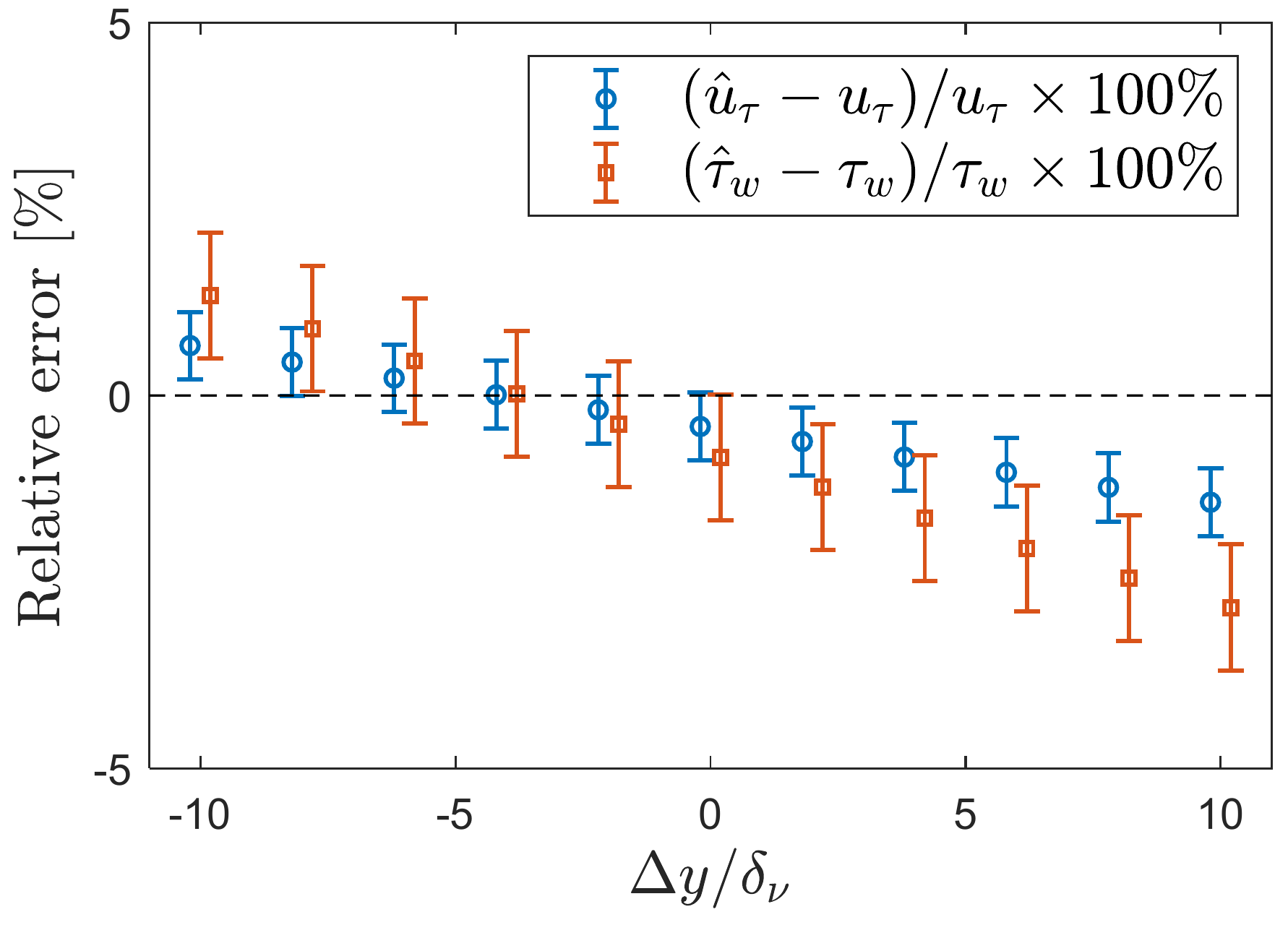}
	\caption{Averaged relative error in the UKF-estimated wall shear stress and friction velocity, when the synthetic PIV measurements are shifted by $\Delta y$ in $y$-direction. The markers in the middle represent averaged relative error. The upper and lower whiskers of the error bars represent the averaged relative uncertainties (i.e., $\pm 2\hat{\sigma}_{u_\tau}/u_\tau\times 100\%$  and $\pm 2\hat{\sigma}_{\tau_w}/\tau_w\times 100\%$) from the UKF.} 
	\label{fig:robust_dy}
\end{figure}

Figure.~\ref{fig:robust_dy} shows relative error in the UKF-estimated wall shear stress and friction velocity when the synthetic PIV measurements are shifted by $\Delta y$ in $y$-direction. 
The markers in the middle of the error bars indicate averaged relative error from  5,000 independent runs of the UKF-based data assimilation algorithm. The upper and lower whisker represent the averaged relative uncertainties for $\hat{u}_\tau$ and $\hat{\tau}$ from UKF (i.e., $\pm 2\hat{\sigma}_{u_\tau}/u_\tau\times 100\%$ in blue, and $\pm 2\hat{\sigma}_{\tau_w}/\tau_w\times 100\%$ in red, respectively). Relative errors in $\hat{u}_\tau$  and $\hat{\tau}_w$ are less than 1\% and  3\%, respectively,  over a wide range of wall offset ($-10 \le \Delta y / \delta_\nu \le 10,$ where $\delta_\nu$ is the viscous length scale).  Particularly, when wall offset $\Delta y = 0$, the averaged relative errors in $\hat{u}_\tau$  and $\hat{\tau}_w$ are about 0.4\% and  0.8\%, respectively, meaning that we expect a finite and small bias from the data assimilation. This small bias is perhaps due to the local inaccuracy of the TBL model, as well as the limited resolution of our experimental data.

In this synthetic flow, $10\delta_\nu$~corresponds to $32.3~{\upmu}$m in physical dimensions. As a reference, a typical $16 \times 16$~pixel interrogation window physical length in a real experiment (e.g., described in \S\ref{sec:exp}) has a length of 0.208 mm, and 75\% interrogation window overlap ratio leads to about 52~${\upmu}$m grid spacing between neighboring PIV vectors. 

\begin{figure}[htbp]
	\centering
	\includegraphics[width=.8\textwidth]{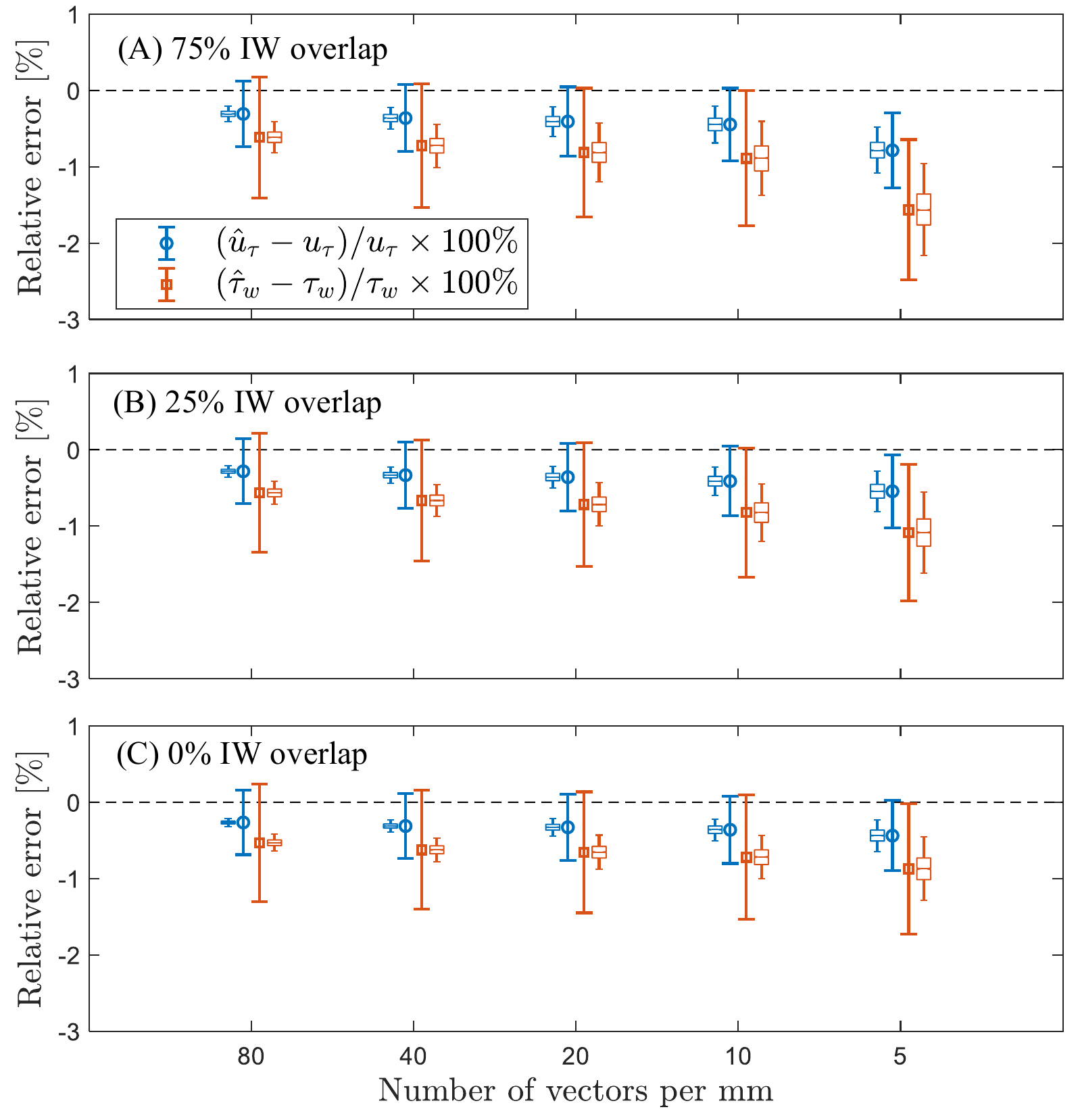}
	\caption{Relative error in the UKF-estimated wall shear stress and friction velocity when the resolution of PIV measurements are varied. (A), (B), and (C) show the results when the interrogation windows (IW) are overlapped by 75\%, 25\%, and 0\%, respectively. The error bars represent the averaged relative error and uncertainty estimates from the UKF, similar to that in Fig. \ref{fig:robust_dy}. The box plots next to the error bars with matching color indicate the scattering of the relative error in the UKF estimates. Each box contains 5,000 independent runs of the UKF -based data assimilation.}
	\label{fig:robust_res}
\end{figure}

Next, we consider the influence of PIV spatial resolution and overlap ratio on the UKF. Figure~\ref{fig:robust_res} shows the relative error of the UKF estimates of wall shear stress and friction velocity. Figure~\ref{fig:robust_res}(A), (B), and (C) show the relative error for PIV data for different interrogation window (IW) overlap ratios, from 75\%, 25\% to 0\%, respectively. 
The blue and red error bars in Fig.~\ref{fig:robust_res} show the averaged relative error and uncertainty of $\hat{u}_\tau$ and $\hat{\tau}_w$, respectively, when spatial resolution of the PIV measurements are varied. 
The markers in the middle indicate the averaged relative error, and the upper and lower whiskers of the bars represent the average uncertainty estimates, all from 5,000 independent runs of the UKF-based data assimilation algorithm.
A wide range of  PIV resolution is tested: from 80~vectors/mm, which is close to the original DNS data resolution in the range of $55<y^+<600$, to 20~vectors/mm, which is similar to the experiments with 75\% overlap shown in \S\ref{fig:exp}, and to much coarser PIV data such as 5~vectors/mm. Even when the PIV resolution is as low as 5~vectors/mm, the relative errors in $\hat{u}_\tau$ and $\hat{\tau}_w$  are less than 1\% and 2\%, respectively. As shown in Fig.~\ref{fig:robust_res}, for each overlap ratio, the markers in the middle of the error bars are closer to the true value (i.e., the horizontal dashed line) for higher PIV resolution cases. This means that the UKF in general provides more \textit{accurate} estimates when higher spatial resolution PIV data are used. 

The box plots to either side of the `error bars' with matching colors indicate the statistical distribution of the relative error in $\hat{\tau}_w$ and $\hat{u}_\tau$ from the 5,000 runs of the UKF. The horizontal bars in the middle of the boxes indicate the median, the upper and lower edge of the boxes are the 25 and 75 percentiles, and the upper and lower whiskers indicate the 2.5 and 97.5 percentiles of the relative error in $\hat{u}_\tau$ and $\hat{\tau}_w$, respectively, and thus represent the 95\% confidence intervals \footnote{ In contrast, the upper and lower whiskers of the (interior) `error bars' indicate the averaged relative uncertainty estimates from the UKF (e.g., $\pm 2\hat{\sigma}_{\tau_w}/\tau_w \times 100\%$).}. As shown in Fig.~\ref{fig:robust_res}, for the same IW overlap ratio, lower PIV resolution leads to taller boxes, meaning that the UKF estimates ($\hat{\tau}_w$ and $\hat{u}_\tau$) are more scattered, namely less \textit{precise}.

Despite the fact that the UKF-based data assimilation algorithm leads to $\hat u_\tau$ and $\hat \tau_w$ that slightly underestimate the true values from the DNS, the algorithm provides accurate estimates (mostly less than 1\% error for $u_\tau$, and less than 2\% error for $\tau_w$) in a large range of the PIV resolution when the interrogation windows (IW) have 75\% overlap (Fig~\ref{fig:robust_res}(A)). One may note that the `error bars' cover the dashed lines in most cases, meaning that the uncertainty estimates of the TBL parameters are statistically valid, except when the PIV resolution is as low as 5~vectors~per~mm. This flawed results can be improved in two ways: i)~increase the spatial resolution of correlated PIV measurement, which is intuitive and can be easily observed in Fig.~\ref{fig:robust_res}(A), (B), or (C);  or ii) reduce the spatial correlation of the low resolution PIV (e.g., reduce the IW overlap ratio) while keeping the same PIV uncertainty. Comparing Fig.~\ref{fig:robust_res}(B) and (C) against (A), we can see that for the same spatial PIV resolution, PIV data from lower IW overlap ratio lead to more accurate and precise estimation results. This improvement is  particularly apparent for low PIV spatial resolution cases. For example, when the PIV resolution is 5 vectors/mm, decreasing IW overlap ratio leads to lower averaged relative error of $\hat{u}_\tau$ and $\hat{\tau}_w$ (marks of the error bars in Fig~\ref{fig:robust_res}(C) and (B) are closer to the horizontal dashed line than that in (A)), as well as shorter boxes, meaning more precise estimates of $\hat{u}_\tau$ and $\hat{\tau}_w$. 

In summary, even 8 vectors (corresponding to 5 vectors/mm in Fig.~\ref{fig:robust_res}(A)) from PIV measurements with 75\% IW overlap ratio can lead to reasonably accurate $\hat{u}_\tau$ and $\hat{\tau}_w$ estimation with approximately 0.5\% and 1\% expected relative error, respectively. The results imply that, in practice, this data assimilation algorithm can be applied to velocimetry techniques with lower spatial resolutions than typical PIV, while higher resolution and lower correlation can enhance the data assimilation performance. 

\section{Experiments and verification}
\label{sec:exp}

\begin{figure}[htpb]
	\centering
	\includegraphics[width=\textwidth]{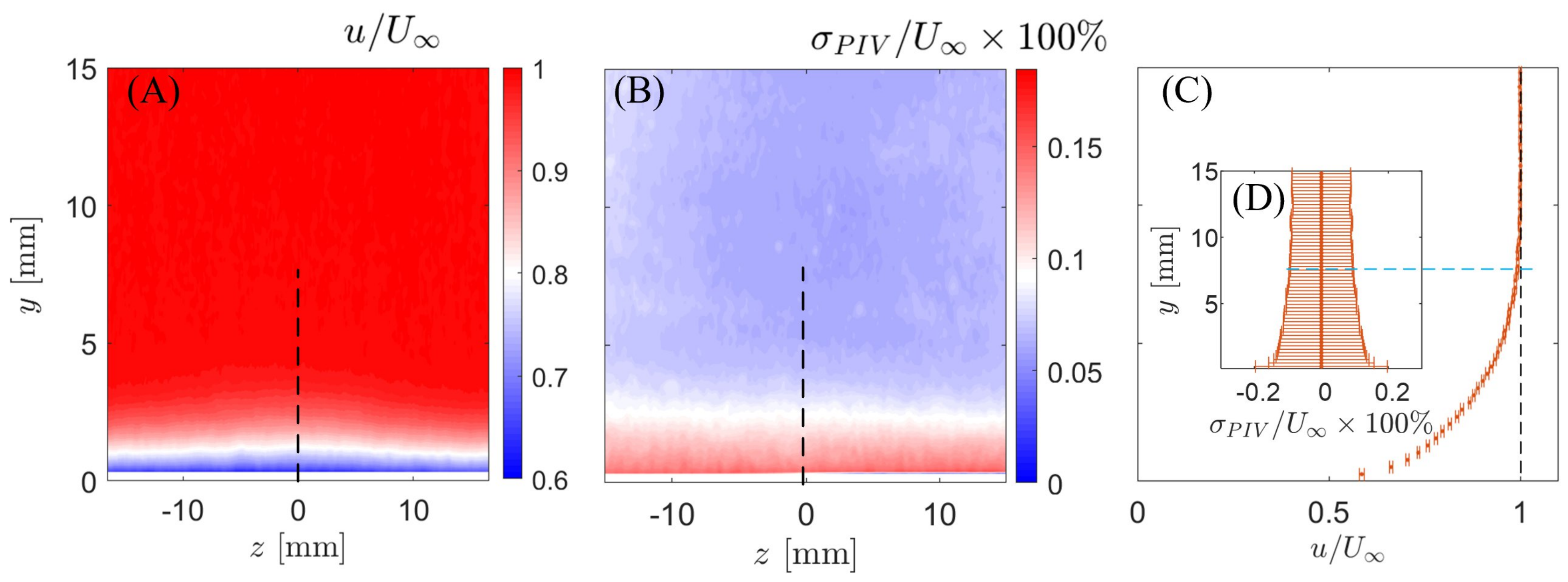}
	\caption{Averaged streamwise velocity profile (A) and uncertainty (B) from $N=500$ snapshots of stereo-PIV ($U_\infty$ is the free stream velocity). The black dashed line indicates where the data (shown in (C)) are sampled and used as input to the UKF algorithm. Velocity profile $u(y)$ and the corresponding uncertainty ($\sigma_{PIV_{i,i}}$) are shown in (C) and (D), respectively. Only 1 out of 4 data points are shown for visualization. The blue dashed line indicates the edge of boundary layer ($\delta_{0.99}$).}
	\label{fig:exp}
\end{figure}

In this section, we apply the UKF-based data assimilation algorithm to the experimental data sets. 
SPIV is used to measure the TBL profile in a channel flow at Mach 0.3. An optical system, consisting of an Evergreen Nd:YAG laser (EVG00400) coupled to a plano-convex lens and a plano-concave cylindrical lens, is used to produce a laser sheet with a thickness of approximately 2 mm at a pulse repetition rate of 10~Hz. Sub-micron (nominally  \SI{0.3} {\micro\metre}) diameter seeding particles are generated using Rosco fog fluid in a customized Wright nebulizer seeder and injected upstream of the stagnation chamber. Two Imager sCMOS cameras (2560~$\times$~2160 pixels, LaVision) are used to acquire the image pairs. Each camera is equipped with a Nikon Micro-Nikkor 105 mm 1:2.8 lens, a Scheimpflug, and a 532 nm band-pass filter. The PIV data and corresponding uncertainty quantification are acquired and processed using DaVis 8.4.0. A multi-pass interrogation window scheme from $48 \times 48$ pixels to $16 \times 16$ pixels with 75\% overlap is applied in the processing, resulting in a resolution of 19 vectors/mm. Universal Outlier Detection \cite{Westerweel2005} is also performed. The uncertainty of each instantaneous vector field is estimated using the correlation statistics method \cite{Wieneke_2015} integrated in DaVis software 8.4.0. In addition, the error of the camera self-calibration process in the vicinity of the extracted velocity profile is estimated to be less than 0.1 pixels and is thus negligible to the other errors \cite{bhattacharya2016stereo}. The mean flow profile is calculated by averaging $N=500$ instantaneous snapshots. The mean flow profile and the uncertainty profile are shown in Fig.~\ref{fig:exp}(A) and (B), respectively. The PIV data used by the data assimilation algorithm is shown in Fig.~\ref{fig:exp}(C).

An alternative method for calculating $\tau_w$ and $u_\tau$ considers the flow in a rectangular control volume that is bounded by the wall and the wind tunnel half-height in the $y$-direction and two SPIV $(y,z)$ planes separated by distance $\Delta L = 19$ cm in the $x$-direction. Applying mass conservation and the $x$-momentum equation leads to the average skin friction coefficient and, consequently, $\tau_w$ and $u_\tau$ as described in \citet{Cantwell2018}. This approach falls in the momentum balance category shown in Fig.~\ref{fig:tbl_measure_class} \cite{winter1979outline}. The resulting $\tau_w$ is not dependent on any assumed flow similarity and thus may produce different results. Monte Carlo simulations are then performed $10^4$ times to estimate the corresponding uncertainty shown in Table~\ref{tab:1}. More details about the control volume method can be found in \citet{Gustavsson2019Acoustic}.

Preston probe measurements are used to provide an independent measurement of friction velocity. Based on the range of validity of the correlation of \citet{ferriss1965preston} and guidelines in \citet{head1962preston}, a probe with the outer diameter of 0.30 mm is chosen. This hypodermic tubing is adhered to the wall with its tip at the streamwise location of an upstream static pressure tap in the test section.
This static tap and the Preston probe are connected to a $5 \pm 0.05$ psid ($35 \pm 0.35$ kPa) differential pressure transducer, PX26-005DV, and the measured differential pressure, $\Delta P$, is used to calculate the local shear stress using Ferriss' correlation based on the data by \citet{head1962preston}. 

Direct wall shear stress measurements are carried out using a flush-mounted, micromachined floating-element sensor from $\text{IC}^2$, Type CS-A05, with a 300 Pa range and a DC measurement accuracy of $\pm 0.3$ Pa. This model has a sensing element size 1~mm~$\times$~0.2~mm and a bandwidth of 5 kHz.

\begin{figure}[htpb]
	\centering
	\includegraphics[width=0.9\textwidth]{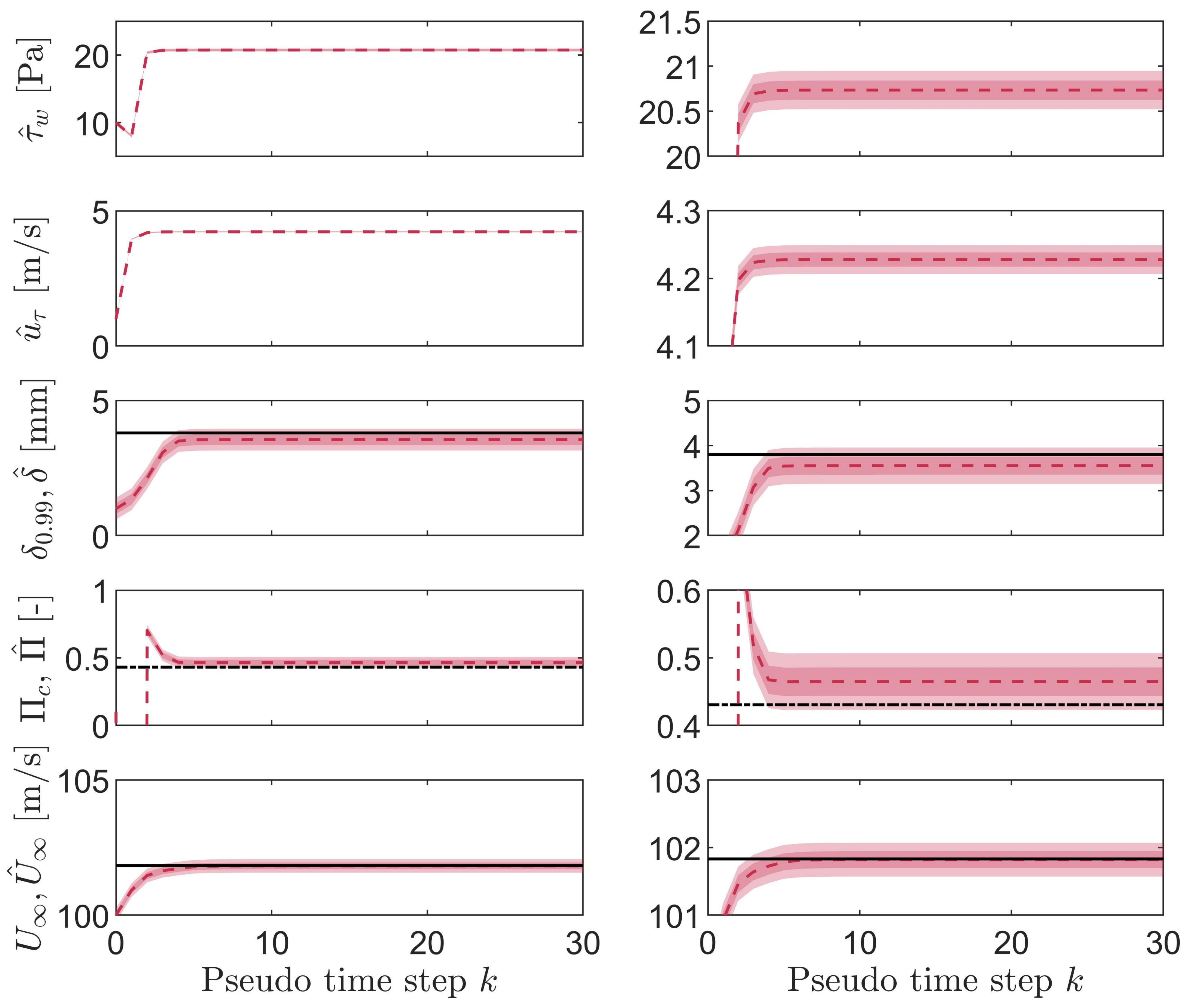}
	\caption{UKF estimation results with experimental measurements. The red dashed lines are the estimated TBL parameters from the UKF. The darker and lighter shades of red indicate the $1\sigma$ and $2\sigma$ uncertainty bands, respectively. The black solid lines indicate the measured values from SPIV data. The black chain line is the calculated~$\Pi$ using the methods in \citet{jones2001evolution}. The corresponding zoomed-in views are shown in the right column.}
	\label{fig:ukf_results_exp}
\end{figure}

Based on the measurements from SPIV, Preston probe, and MEMS wall shear stress sensor, the UKF-based data assimilation is performed and the result is shown in Fig.~\ref{fig:ukf_results_exp}. The UKF converges in a few steps and the estimates compare favorably with measurable values (e.g., $\delta_{0.99}$ and $U_\infty$ that are directly measured from PIV data), and measurements from other independent techniques (see also Table~\ref{tab:1}).

\begin{figure}[htpb]
	\centering
	\includegraphics[width=0.55\textwidth]{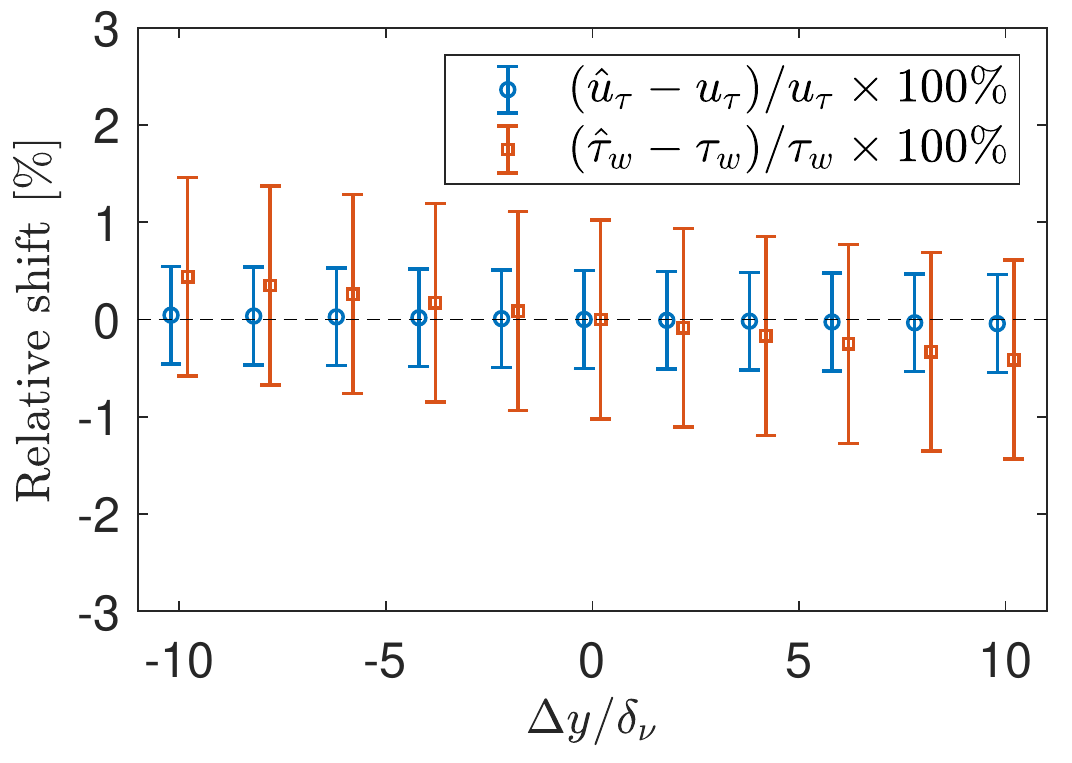}
	\caption{Relative shift in the UKF-estimated wall shear stress and friction velocity, when the SPIV measurements are shifted by $\Delta y$ in $y$-direction. The markers in the middle represent relative error when the wall location is not shifted. The upper and lower whiskers of the error bars represent the averaged relative uncertainties (i.e., $\pm 2\hat{\sigma}_{u_\tau}/u_\tau\times 100\%$  and $\pm 2\hat{\sigma}_{\tau_w}/\tau_w\times 100\%$) from the UKF.}
	\label{fig:robustness_dy_exp}
\end{figure}

In order to test the robustness of data assimilation algorithm to wall offset, we perturb the wall location by $-10 \le \Delta y / \delta_\nu \le 10$, where $10\delta_\nu$~corresponds to $51.1~{\upmu}$m in physical dimensions, which is close to grid spacing between neighboring PIV vectors. As shown in Fig.~\ref{fig:robustness_dy_exp}, the expected value of relative shift in the estimated friction velocity ($\hat u_{\tau}$) and wall shear stress ($\hat \tau_w$) is less than 0.1\% and 0.5\%, respectively. Note, the unperturbed estimates from the UKF are used as reference value. This result based on experimental data shows again that the UKF-based data assimilation algorithm is robust to potential inaccurate wall position measurement.

\begin{figure}[htbp]
	\centering
	\includegraphics[width=0.9\textwidth]{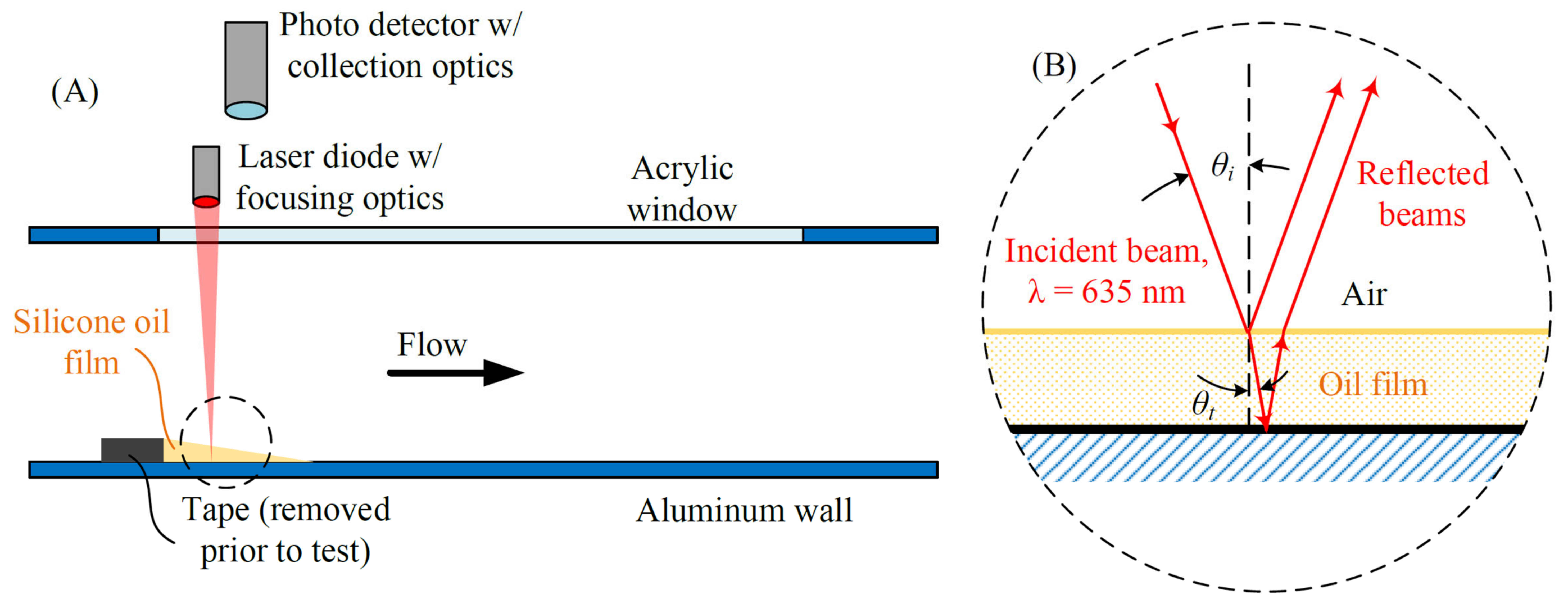}
	\caption{Schematic of laser interferometer skin friction meter setup (A), and the corresponding optical path (B).}
	\label{fig:lisf_setup}
\end{figure}

To further validate $u_\tau$ and $\tau_w$ obtained from the methods described previously, Laser Interferometer Skin Friction (LISF) measurements are carried out, as shown in Fig.~\ref{fig:lisf_setup} using the methodology described in \citet{Garrison:2012ic} for high-speed flows.  In this technique, a thin layer of oil with known viscosity determined as a function of temperature is applied on the polished wall. The wall shear stress induced by the flow leads to a thinning of the oil film over time. If a laser beam impinges the oil-covered wall surface, two reflections are obtained, one from the top of the oil film and one from the wall surface itself. When the oil film thickness reduces to a few times that of the wavelength of the laser light, oscillations appear in the photo detector signal due to constructive and destructive interference between the reflected beams. Further details of the LISF technique are given by \citet{Garrison1994}.

The current implementation used a 635 nm laser diode with a lens focusing the beam on the surface of a flat plate mounted in the bottom of the wind tunnel test section. The reflected signal is measured by a Thorlabs DET10A photodetector fitted with a 25~mm lens for light collection and mounted next to the laser diode. Dow Corning silicone oil with a nominal viscosity of 100~cSt is used. The temperature of the oil is estimated based on a recovery temperature obtained from the measured Mach number and stagnation temperature together with a turbulent boundary layer recovery factor of~0.89. As shown in Fig.~\ref{fig:lisf_setup}, a piece of tape is used to define the upstream edge of the oil as it is applied. Excess oil is then scraped off, producing a sub-mm thick wedge of oil downstream of the tape. The tape is then removed before the test is executed. The data reduction procedure, including the estimation of measurement uncertainty, is provided in \ref{appB}.

With the  experimental data and the corresponding uncertainty from the SPIV, Preston tube, and the shear stress sensor, the UKF-based data assimilation algorithm provides optimal estimates of $u_\tau$ and $\tau_w$ and their nondimensional equivalent values (Table~\ref{tab:1}). In addition to comparing the results with measured values from shear stress sensor, Preston tube, control volume method, and LISF, we also estimated $u_\tau$ via the Clauser method \citep{Clauser:1956vl}. The values of $u_\tau$ and $\tau_w$ are slightly dependent on the range of the data used in the fit (Table~\ref{tab:1}). Most of the measurement techniques give similar estimates of $u_\tau$ and $\tau_w$, except for the wall shear stress sensor, which is expected to be biased due to the pressure gradient \cite{Mark}. Note LISF provides a result that is less than the other results. However, considering the different freestream conditions (Mach 0.286 in the LISF experiments versus Mach 0.295-0.311 in the other experiments) and the uncertainty in the LISF technique, the discrepancy is not unreasonable.

\begin{table}
		\centering
	\caption{Cross validation of the measured and estimated TBL parameters.}
	\label{tab:1}
		\begin{threeparttable}
					\begin{tabular}{ccccc}
			\toprule 
			& $u_\tau$ [m/s] & $\tau_w$ [Pa]  & $S$ [-] & $C_f$ [$\times10^{-3}$] \\ 
			\midrule
			Shear stress sensor$^1$ 		 & $3.3 \pm 0.8$          & $13 \pm 10$  & $32 \pm 8$ & $2 \pm 1$      \\
			Preston tube        		 & $4.3 \pm 0.3$          & $21.5 \pm 1.7$  &  23.8 $\pm$ 1.4 & 3.5 $\pm$ 0.2     \\
			UKF                  		 & 4.23 $\pm$ 0.02   & 20.73 $\pm$ 0.21  & 24.09 $\pm$ 0.12  & 3.45 $\pm$ 0.04 \\
			Clauser method$^2$ 	 & 4.28 $\pm$ 0.07     & 20.99 $\pm$ 0.68   & 24.17$\pm$ 0.55 &  3.59 $\pm$ 0.11 \\
			Control volume       		 & 4.16 $\pm$ 0.18     & 20.23 $\pm$ 1.76 & 25.11 $\pm$ 1.10 & 3.24 $\pm$ 0.31 \\ 
			LISF             & 3.83 [3.63 3.94]  & 17.3 $[15.5, 18.3]$  & 25.5 [24.8 26.8]  & 3.1 $[2.7, 3.3]$\\ 
			\bottomrule
				\end{tabular}
		\begin{tablenotes}
			\footnotesize
			\item $^1$The estimated uncertainty of the shear stress sensor here is the uncertainty used in the data assimilation algorithm.
			\item $^2$Values vary depending on range of data used in fitting.
			\end{tablenotes}
	\end{threeparttable}
\end{table}

\section{Conclusions and discussion}
\label{sec:conclusions}
The results indicate that the UKF-based TBL parameter estimation is robust and able to produce accurate estimates with quantifiable uncertainty. In addition, the approach naturally handles the four challenges mentioned in \S\ref{sec:introduction}.  The UKF-based data assimilation is a useful framework that can fuse multiple measurements and output robust optimal estimates of TBL parameters. In contrast to a nonlinear regression method, which seeks parameters that minimize a chosen error norm, the UKF-based data assimilation method minimizes the covariance of the estimates, thereby producing parameter estimates with quantifiable uncertainty. The data assimilation algorithm directly leverages uncertainty information of available experimental measurements. It is worth noting that the uncertainty or noise properties of the UKF are evaluated via covariance matrices, which naturally take higher-order statistical information of measurements (e.g., PIV uncertainties are correlated) into account. By incorporating this information, the UKF systematically and objectively tolerates inaccurate measurements (e.g., shear stress sensor in the validation example in \S\ref{sec:validation}) and leverage measurements with lower uncertainty (e.g., SPIV in the validation example in \S\ref{sec:validation}).

The UKF-based data assimilation algorithm is also flexible. In the present studies, measurements and corresponding uncertainties of only PIV, Preston tube, and shear stress sensor are used and independently validated using a control volume analysis and LISF. This is just one example application of the UKF-based data assimilation framework. In fact, combinations of other measurements are also possible to design new data assimilation algorithms for TBL parameter estimation. We leave these options for future work.

We propose that a data assimilation algorithm (e.g., the UKF-based data assimilation introduced in the current work) should be used instead of nonlinear regression in practical TBL parameter estimation. Such a data assimilation framework can effectively combine the appropriate physical model (such as the modified Musker's profile for zero pressure gradient TBL) with available experimental measurements to provide robust estimates with quantifiable uncertainty. The UKF-based method offers the advantage of \textit{explicitly} handling uncertainty propagation from measurements to estimates and can leverage future advances in measurement techniques and their uncertainty quantification.


At last, it should be noted that the data assimilation algorithm in the current works is designed by embedding a semi-analytical model of canonical TBL \cite{musker1979explicit,rodriguez2015robust}, which is able to accurately describe the flow profile of a zero-pressure-gradient TBL. To apply the similar data assimilation algorithm to TBL with adverse or favorable pressure gradients, we expect that a flow profile model that accounts for non-zero pressure gradient could be used for accurate estimation \cite{indinger2006mean,durbin1992scaling}. We will leave this topics for future studies.

\section{Acknowledgments}
The computational time for the direct numerical simulations was supported by SciNet  (www.scinethpc.ca) and Compute Canada (www.computecanada.ca).

\appendix 
\section{Unscented transform and the unscented Kalman filter}
\label{appA0}
This appendix section provides implementation details of the unscented Kalman filter~(UKF), which utilizes the unscented transform~(UT) to propagate the random variable through a nonlinear process. Despite the theory and applications of the UKF are well-documented in, for example, \citet{wan2000unscented,julier2004unscented,simon2006optimal}, a brief summary of the UT and UKF are provided here for convenience.

\subsection{Unscented Transform}
Consider a nonlinear function $\bm{Y} = \text{fun}(\bm{X})$ that maps a $L$-dimensional random variable from $\bm{X}$~to~$\bm{Y}$. Assuming the first two moments of $\bm{X}$ are $\E[\bm{X}]$ and $\Cov[\bm{X}]$, respectively, the statistics of $\bm{Y}$ can be calculated via the following two steps:

Step 1, generate $2L+1$ weighted sigma points $\bm{\mathcal{\bm{X}}}_i$ around $\bm{X}_i$: 
\begin{equation} 
\label{eq:UTSigma}
\begin{split}
 \bm{\mathcal{X}}_1 & = \E[\bm{X}]\\
 \bm{\mathcal{X}}_i & = \E[\bm{X}] + \left( \sqrt{(L+ \lambda)\Cov[\bm{X}]} \right)_i \quad  i = 2, \dots, L+1\\
  \bm{\mathcal{X}}_i & = \E[\bm{X}] - \left( \sqrt{(L+ \lambda)\Cov[\bm{X}]} \right)_i \quad  i = L+2, \dots, 2L+1
\end{split}
\end{equation}
and the corresponding weight vector for expected value and covariance calculation are
\begin{equation} 
\label{eq:UTweight}
\begin{split}
\bm{W}_1^{E} & = \frac{\lambda}{\lambda + L}\\
\bm{W}_1^{Cov} & = \frac{\lambda}{\lambda + L} + (1 - \alpha^2 + \beta)\\
\bm{W}_i^{E}  = \bm{W}_i^{Cov} &= \frac{1}{2(\lambda + L)} \quad  i = 2, \dots, 2L+1, 
\end{split}
\end{equation}
where subscript $(\cdot)_i$ indicates the $i$-th row of the corresponding matrix or column vector. In \eqref{eq:UTSigma} and \eqref{eq:UTweight}, $\lambda = (\alpha^2-1)L,$ where $\alpha$ is typically a small number that controls the scatting of the sigma points ($\alpha = 0.01$ is used in the present work\footnote{A wide range  of $\alpha$ is tested ($\alpha = 10^{-5}$~--~$10^{-1}$), and no significant effects are found to the UKF performance in the present work.}), and $\beta =2$ is set to best cooperate with Gaussian variables \cite{wan2000unscented}. The corresponding sigma points of $\bm{Y}$ can be directly calculated by propagating sigma points~$\bm{\mathcal{X}}$ through the nonlinear mapping, for example, $\bm{\mathcal{Y}}_i = \mathcal{H}(\bm{\mathcal{X}}_i)$.  

Step 2, the mean and covariance of $\bm{Y}$ can be approximated as
\begin{equation} 
\label{eq:UT E and Cov}
\begin{split}
 \Bar{\bm{Y}} = \E[\bm{Y}] & \approx  \bm{\mathcal{Y}} \bm{W}^{E}  \\
\bm{P} = \Cov[\bm{Y}] & \approx (\bm{\mathcal{Y}} - \E[\bm{Y}]\bm{J}_{1,m}) \textbf{diag}\left(\bm{W}^{Cov}\right) (\bm{\mathcal{Y}} - \E[\bm{Y}]\bm{J}_{1,m})^T, \\
\end{split}
\end{equation}
respectively, where $\bm{J}_{1,m}$ is an $1 \times m$ all-one matrix ($m=2L+1$). 

\subsection{Unscented Kalman Filter}
The UKF algorithm in the present research is implemented based on \citet{wan2000unscented} under the classic Kalman filter framework \cite{simon2006optimal}:

1. Initialize UKF with some guesses of the $\hat{\bm{X}}_0$ and $\hat{\bm{P}}_0$, for example:  
\begin{align*} 
\hat{\bm{X}}_0  = \bm{X}_0 \\ 
\hat{\bm{P}}_0  =  \bm{Q} .
\end{align*}

2. For $k = 1, \dots, \infty,$  choose sigma points
$\bm{\mathcal{X}}_k = \left[  \hat{\bm{X}}_{k-1}, \hat{\bm{X}}_{k-1}\pm \sqrt{(L+\lambda) \hat{\bm{P}}_{k-1}},   \right]$
recursively and predict \textit{a priori} estimates, denoted with superscript $(\cdot)^-$, as well as the corresponding covariance:  
\begin{align*} 
\bm{\mathcal{X}}_{k|k-1} &= \mathcal{F}\left(\bm{\mathcal{X}}_{k-1}\right)\\
\hat{\bm{X}}_k^- &=\bm{\mathcal{X}}_{k|k-1} \bm{W}^E\\
\hat{\bm{P}}_k^- &= (\bm{\mathcal{X}}_{k|k-1} - \hat{\bm{X}}_k^-\bm{J}_{1,m})\textbf{diag}\left(\bm{W}^{Cov}\right)(\bm{\mathcal{X}}_{k|k-1} - \hat{\bm{X}}_k^-\bm{J}_{1,m})^T + \bm{Q}\\
\bm{\mathcal{Y}}_{k|k-1} &= \mathcal{H}(\bm{\mathcal{X}}_{k|k-1})\\
\hat{\bm{Y}}_k^- &= \bm{\mathcal{Y}}_{k|k-1} \bm{W}^E\\
\bm{P}_{Y_k,Y_k} &= (\bm{\mathcal{Y}}_{k|k-1} - \hat{\bm{Y}}_k^-\bm{J}_{1,m})\textbf{diag} \left(\bm{W}^{Cov}\right)(\bm{\mathcal{Y}}_{k|k-1} - \hat{\bm{Y}}_k^-\bm{J}_{1,m})^T + \bm{R}\\
\bm{P}_{X_k,Y_k} &= (\bm{\mathcal{X}}_{k|k-1} - \hat{\bm{X}}_k^-\bm{J}_{1,m})\textbf{diag} \left(\bm{W}^{Cov}\right)(\bm{\mathcal{Y}}_{k|k-1} - \hat{\bm{Y}}_k^-\bm{J}_{1,m})^T, 
\end{align*} 
where $\bm{Q}$ and $\bm{R}$ are covariance matrices of process and observation noise, respectively.   

3. Correct the predictions using measurements ($\tilde{\bm{Y}}_k$) with a Kalman gain $\bm{\mathcal{K}}_k$ to provide \textit{a posteriori} estimates, denoted with accent $(\hat \cdot)$: 
\begin{align*} 
\bm{\mathcal{K}}_k &= \bm{P}_{X_k,Y_k} \bm{P}_{Y_k,Y_k}^{-1}\\
\hat{\bm{X}}_k &= \hat{\bm{X}}_k^- +  \bm{\mathcal{K}}_k( \tilde{\bm{Y}}_k - \hat{\bm{Y}}_k^- )\\
\hat{\bm{P}}_k &= \hat{\bm{P}}^-_k  -   \bm{\mathcal{K}}_k \bm{P}_{Y_k,Y_k}  \bm{\mathcal{K}}_k^T,
\end{align*}
until $\hat{\bm{X}}_k$ and  $\hat{\bm{P}}_k $ are converged.

\section{LISF data reduction procedure and measurement uncertainty}
\label{appB}
\begin{figure}[!htpb]
	\centering
	\includegraphics[width=0.5\textwidth]{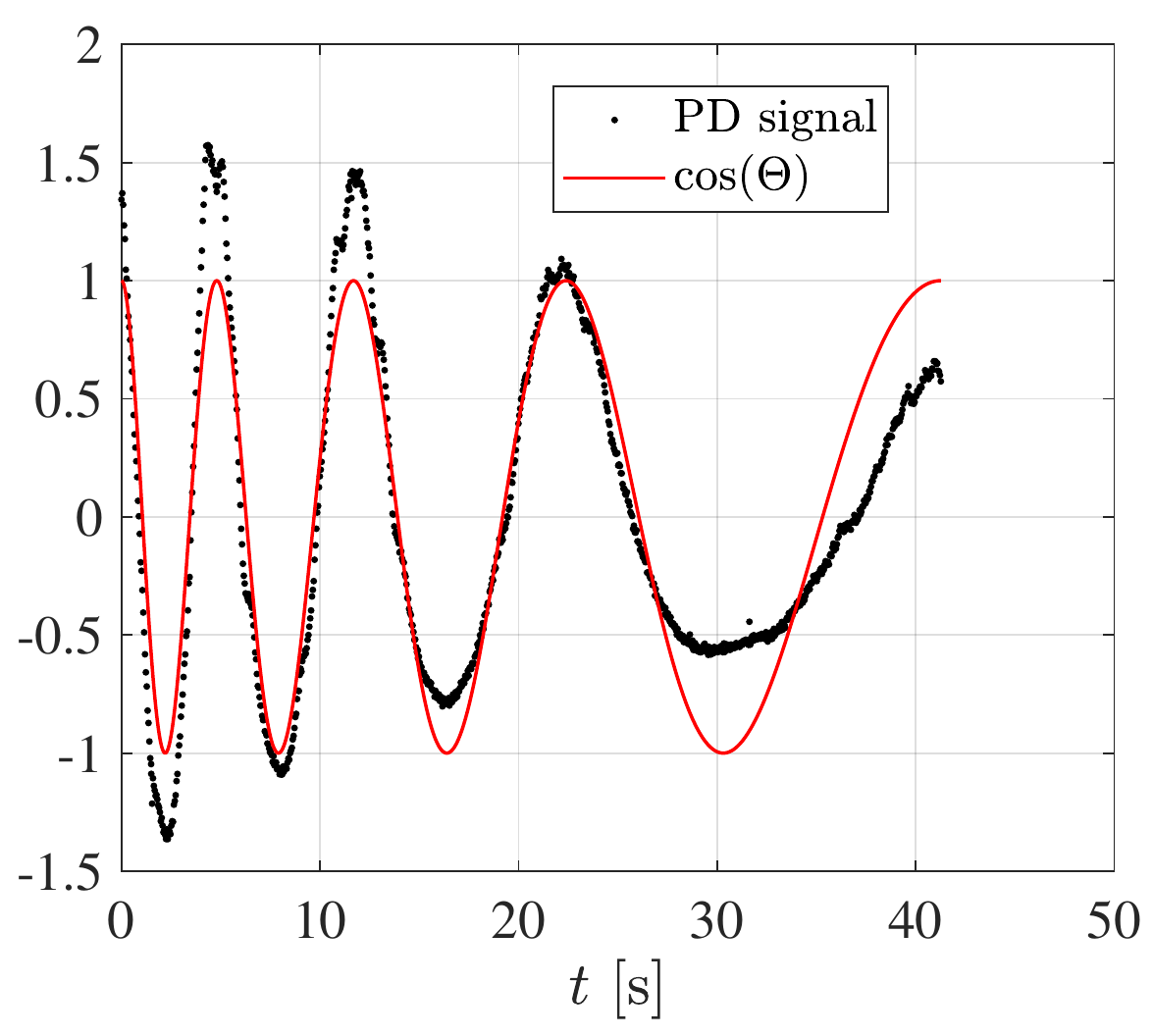}
	\caption{Data reduction results for the LISF experiment. The scaled and detrended photodetector signal is compared to the theoretical $\cos(\Theta)$ function of the phase difference $\Theta$ between the two laser beams. The wall shear stress corresponds to the value that provides the best temporal alignment of the peaks and valleys of the waveforms.}
	\label{fig:data_reduction}
\end{figure}

The photodiode signal acquired using the setup shown in Fig.~\ref{fig:lisf_setup} is processed via the method employed by \citet{Garrison1994}. As the shear stress of the flow acts on the oil film, it thins over time, reducing the phase difference $\Theta$ between the beam reflected from the top surface of the oil film and the beam reflected from the wall. Initially, the shear stress produces short-wavelength surface ripples that contaminate the signal, but once the film has grown thin enough, the signal becomes usable, varying as a cosine function with a frequency that decreases with time. Manual inspection is used to identify this time interval, as shown in Fig.~\ref{fig:data_reduction}.  This representative signal starts with a local maximum associated with constructive interference and ends with another local maximum 4 periods or approximately 41~sec later. 

The data reduction method uses the \citet{Garrison1994} model that predicts the theoretical oil film interference signal during this time interval as a function of the wall shear stress $\tau_w$. The PD signal is detrended to remove the mean and arbitrarily scaled in amplitude, and then the correlation coefficient is computed to find the best match of the peaks and valleys between the signals.  The matched theoretical $\cos(\Theta)$ signal is shown as a solid red line in the figure. Note that the amplitude of the signals are not important, only the temporal alignment of the local maxima and minima.

The primary contributions to the uncertainty in the estimated wall shear stress are the start and end points of the useful signal range, the stagnation and wall temperatures (which determines the oil viscosity), the Mach number, the laser beam incidence angle, and the distance between the leading edge of the oil film and the laser spot.  The influence of the estimated uncertainty in these parameters is assessed using a set of 10,000 Monte Carlo simulations. In particular, the starting time (associated with the sharp first peak) is perturbed by just $\pm$~2 time steps while the end time (associated with the last broad peak) is perturbed by $\pm$~15 time steps. The locations of the lower 2.5\% and upper 97.5\% probability levels provide the lower and upper 95\% confidence intervals shown in brackets in Table~\ref{tab:1}.

\bibliographystyle{dcu}
\bibliography{ISPIV2019_References}
\end{document}